\documentclass[11pt,a4paper]{article}

\usepackage{amsmath,amssymb,amsfonts}
\usepackage{graphicx}
\usepackage{float}
\usepackage{geometry}
\usepackage{setspace}
\usepackage{caption}
\usepackage{hyperref}
\usepackage{xcolor}
\usepackage{comment}

\newcommand{\be}{\begin{equation}}
\newcommand{\ee}{\end{equation}}
\newcommand{\bea}{\begin{eqnarray}}
\newcommand{\eea}{\end{eqnarray}}

\newcommand{\sch}{Schwarzschild}

\newcommand{\TS}{Tomimatsu--Sato}

\numberwithin{equation}{section}
\numberwithin{figure}{section}

\begin{document}
	
	\begin{titlepage}
		\vspace*{1cm}
		\begin{center}
			{\Large \bf Tomimatsu--Sato $\delta=2$ Solution in Kaluza--Klein Theory}
		\end{center}
		\vspace{1cm}
		\begin{center}
			\renewcommand{\thefootnote}{\fnsymbol{footnote}}
			Haryanto M.~Siahaan$^{1}$\footnote{haryanto.siahaan@unpar.ac.id},
			Masoud Ghezelbash$^{2}$\footnote{amg142@campus.usask.ca}
			\\[1cm]
			$^{1}$Department of Physics, Parahyangan Catholic University,\\
			Jalan Ciumbuleuit 94, Bandung 40141, Indonesia
			\\[0.5cm]
			$^{2}$Department of Physics and Engineering Physics,\\
			University of Saskatchewan, Saskatoon, Saskatchewan S7N 5E2, Canada
			\renewcommand{\thefootnote}{\arabic{footnote}}
		\end{center}
		\vspace{1cm}
		\begin{abstract}
We construct the Kaluza-Klein electric dressing of the Tomimatsu-Sato $\delta=2$ vacuum spacetime by uplifting to five dimensions, boosting along the compact direction and reducing to Einstein-Maxwell-dilaton theory with coupling $a=\sqrt3$. It carries a nontrivial dilaton, unlike the known Einstein-Maxwell charged and magnetized Tomimatsu-Sato solutions. At leading order it shares the mass-charge-dilaton relations and gyromagnetic ratio $g=2-\tanh^2\alpha$ of the boosted Kerr black hole; the seed first shows up in the quadrupole. Using Mayerson's non-vacuum Geroch-Hansen multipoles, we derive the exact law $M_2(\alpha)=\frac12(1+\cosh^2\alpha)M_2^{(0)}$, so the $\delta=2$ quadrupole excess $\Delta M_2=-\frac14p^2m^2M$ survives every boost. In the Kodama-Hikida picture the Weyl segment $\rho=0$, $|z|<\kappa$ is a singular rod with degenerate Killing horizons at its endpoints. There the boost function equals $\cosh\alpha$, so the near-horizon geometry is a constant rescaling of the seed's: the horizons stay degenerate and non-rotating, their areas and Komar angular momenta scale by $\cosh\alpha$, and the electric potential is $\frac12\tanh\alpha$. The electric flux through any closed surface is $\sinh\alpha\cosh\alpha$ times the enclosed seed Komar mass, so the horizons and the ring singularity are uncharged and the rod carries all the charge; the boost-invariant identity $M-\Phi_eQ=m$ follows. The ergosurface stays at $f=0$; a new curvature-singular boundary appears at $f=\coth^2\alpha$, where the Kaluza-Klein circle becomes null. As the seed potential $f$ is unbounded above near the ring, it exists for every nonzero boost: unlike for Kerr, no Kaluza-Klein boost of TS$_2$ is a spacelike reduction everywhere. The excised region is compact, has the ring on its boundary and never reaches the horizons. Charging pointwise strengthens the causality violation near the rod.
		\end{abstract}
	\end{titlepage}
	
	\section{Introduction}
	\label{sec:intro}
	
The \TS{} (TS) family \cite{Tomimatsu:1972zz,Tomimatsu:1973gq} is the canonical
stationary generalization of the Kerr solution: a discrete deformation parameter
$\delta$ interpolates between Kerr ($\delta=1$) and vacuum spacetimes with
increasingly rich multipole structure, of which $\delta=2$ (TS$_2$) is the
simplest representative. Its global structure is intricate. Early work
identified a naked ring singularity and a region of causality violation
\cite{Gibbons:1973rb}; Kodama and Hikida then showed that, contrary to the
long-held belief that even-$\delta$ TS solutions possess no horizon, TS$_2$ has
a \emph{degenerate} Killing horizon with two components, located at the two
endpoints of the Weyl segment $\rho=0$, $|z|<\kappa$, while the segment itself
is a singular rod with closed timelike curves (CTCs) in its vicinity
\cite{Kodama:2003ch}. Gegenberg \emph{et al.}\ later showed that the rod
carries all of the Komar mass, that the two extremal horizons carry angular
momentum but no mass, and that their near-horizon geometry admits a Kerr/CFT
description \cite{Gegenberg:2010yx}; see also \cite{Manko:2012ts} for a
reassessment of the physical interpretation of the seed. The TS$_2$ spacetime
therefore combines, in exact closed form, essentially all the exotic features
that distinguish non-Kerr geometries.

Charged generalizations of TS$_2$ have so far been constructed within
Einstein--Maxwell theory. Ernst obtained a charged version of the TS
spinning-mass field \cite{Ernst:1973ts}, and Manko, Manko, and
Sanabria-G\'omez constructed a four-parameter charged and magnetized TS$_2$
solution \cite{Manko:1998ts,Manko:1999ts}, whose multipole structure has
recently been analysed with the electrovac moment formalism
\cite{Fodor:2021em}; a striking feature of that solution is that a
sufficiently strong magnetic field removes the ring singularity in the
subextreme regime \cite{Manko:1999ts}. None of these solutions carries a
scalar field.

A different charging mechanism is available in Kaluza--Klein (KK) theory
\cite{Kaluza:1921tu,Klein:1926tv}. Reduction of five-dimensional vacuum gravity
on a circle yields four-dimensional Einstein--Maxwell--dilaton (EMD) theory
with the specific coupling $a=\sqrt3$, and stationary solutions are generated
from four-dimensional vacuum seeds by the uplift--boost--reduce procedure
\cite{Chodos:1981nm,Dobiasch:1981vh,Gibbons:1985ac,Frolov:1987rj}. Applied to
the \sch{} and Kerr seeds it produces the static and rotating KK black holes
\cite{Gibbons:1985ac,Frolov:1987rj,Rasheed:1995zv,Larsen:1999pp,Aliev:2008wv},
and it has been applied systematically to vacuum seeds with KK asymptotics
\cite{Knoll:2015nva}; the group-theoretic structure behind it is reviewed in
\cite{Breitenlohner:1987dg}. The same construction underlies the charged
dilatonic Zipoy--Voorhees solutions \cite{Yazadjiev:2013qna,Yazadjiev:2013hxa,
Siahaan:2026zvd}, a number of magnetized and accelerating KK black holes
\cite{Gibbons:2013dna,Siahaan:2024ilq,Siahaan:2025mkk}, and a broader
programme of exact EMD solutions \cite{Ghezelbash:2017emd,Fahim:2021emd}. The
charged dilatonic Zipoy--Voorhees $\delta=2$ metric is precisely the static
limit of the solution constructed here (Sec.~\ref{subsec:static}).

The problem addressed in this paper is thus well posed: what does the KK boost
do to the TS$_2$ geometry, whose multipolar and causal structure is far richer
than Kerr's, and which already possesses degenerate horizons, a singular rod,
and a ring singularity? The construction itself is standard, and the resulting
fields solve the EMD equations exactly (we verified all components of the
field equations symbolically). What is new is the physics: the KK dressing of
TS$_2$ carries a nontrivial dilaton, its electric and dilaton charges are tied
to a single boost parameter, and --- unlike the Einstein--Maxwell dressings ---
it leaves the ring singularity in place while generating an additional
curvature-singular boundary where the KK circle degenerates. Our main results
are the following.

\begin{enumerate}
\item At leading asymptotic order the mass, angular momentum, electric and
dilaton charges, magnetic dipole, KK constraint, and gyromagnetic ratio
$g=2-\tanh^{2}\!\alpha$ coincide with those of the boosted Kerr black hole
\cite{Frolov:1987rj,Rasheed:1995zv}; the $\delta=2$ character is invisible
there.
\item The seed reveals itself at quadrupole order. We recover the
Geroch--Hansen moments of the vacuum seed, $M_{2}=-m^{3}(q^{2}+p^{2}/4)$
\cite{Tomimatsu:1973gq,Manko:2012ts} and $S_{3}=-m^{4}q(q^{2}+p^{2}/2)$
\cite{Manko:1999ts}, and compute, for the charged solution, the invariant
quadrupole defined through an asymptotically Cartesian mass-centred (ACMC)
chart \cite{Thorne:1980ru}, which Mayerson's generalization of the
Geroch--Hansen formalism \cite{Mayerson:2022ekj} makes well defined for this
non-vacuum spacetime. We obtain, symbolically in $p$ and $\sinh^{2}\!\alpha$,
the exact boost-scaling law $M_{2}(\alpha)=(M/m)\,M_{2}^{(0)}$ for both the
TS$_2$ and the Kerr seeds (Appendix~\ref{app:M2}); the $\delta=2$ excess
$\Delta M_{2}=-\tfrac14p^{2}m^{2}M$ survives at every boost.
\item On the two degenerate horizons the boost function is exactly
$H=\cosh\alpha$, and the near-horizon geometry of the charged solution is a
constant rescaling of the seed's. The horizons stay degenerate and
non-rotating, their areas and Komar angular momenta scale by $\cosh\alpha$,
and the electric potential on them is the constant $\tfrac12\tanh\alpha$. The
electric flux through any closed surface is $\sinh\alpha\cosh\alpha$ times the
seed Komar mass it encloses, an exact consequence of the five-dimensional
Komar identity; hence the horizons and the ring singularity carry no charge,
the entire charge is sourced by the singular rod, and the global identity
$M-\Phi_{e}Q=m$ is boost-invariant.
\item The ergosurface is pinned at the seed locus $f=0$. A new
curvature-singular boundary appears on $f=\coth^{2}\!\alpha$, where the KK
circle becomes null. A boost is an admissible spacelike reduction everywhere
only if $f<\coth^2\!\alpha$ throughout the seed; this holds for every boost of
Kerr but, because $f$ is unbounded above near the TS$_2$ ring, for no boost of
TS$_2$, so the new boundary exists for every nonzero boost throughout the
rotating family $0<p<1$. The excised region is compact, has the ring on its
boundary, and never reaches the horizons; it shrinks onto the ring as
$\alpha\to0$ and approaches the region $f\ge1$ (whose boundary meets the rod at
its equator) as $\alpha\to\infty$. Charging strengthens the
causality violation adjacent to the rod pointwise throughout the physical
domain.
\end{enumerate}

The paper is organized as follows. Section~\ref{sec:KK} reviews the KK
reduction and the boost. Section~\ref{sec:TS} presents the seed, its charges,
and its Geroch--Hansen moments. Section~\ref{sec:charged_TS} assembles the
charged solution and its static limit. Section~\ref{sec:properties} contains
the conserved charges, the quadrupole law, the rod and horizon analysis, and
the global structure. Section~\ref{sec:conclusions} summarizes.
Appendix~\ref{app:M2} derives the quadrupole law.
	
	\section{Kaluza--Klein Reduction and Boosting}
	\label{sec:KK}
	
	\subsection{Effective action and field equations}
	
	We distinguish between the vacuum seed metric and the reduced four-dimensional
	KK fields by writing them as $\tilde g_{\mu\nu}$ and
	$\{g_{\mu\nu},A_\mu,\Phi\}$, respectively. The latter solve the standard
	four-dimensional KK Einstein--Maxwell--dilaton theory, whose action may be
	written as~\cite{Aliev:2008wv}
	\be \label{eq:action}
	S = \int d^4x\,\sqrt{-g}\,\Bigl[ R - 2\,\partial_\mu\Phi\,\partial^\mu\Phi
	- e^{2\sqrt{3}\,\Phi}\,F_{\mu\nu}F^{\mu\nu} \Bigr],
	\ee
	where $R$ is the Ricci scalar of $g_{\mu\nu}$, $\Phi$ is the dilaton,
	and $F_{\mu\nu}=\partial_\mu A_\nu-\partial_\nu A_\mu$.
	
	Variation of~\eqref{eq:action} yields
	\be\label{eq:Rmn}
	R_{\mu\nu}=2\,\partial_\mu\Phi\,\partial_\nu\Phi
	+2\,e^{2\sqrt{3}\,\Phi}\Bigl[F_{\mu\alpha}F_{\nu}^{\ \alpha}
	-\tfrac14 g_{\mu\nu}F_{\alpha\beta}F^{\alpha\beta}\Bigr],
	\ee
	\be\label{eq:Am}
	\nabla_\mu\Bigl(e^{2\sqrt{3}\,\Phi}F^{\mu\nu}\Bigr)=0,
	\ee
	\be \label{eq:Phi}
	2\,\nabla^2\Phi-\sqrt{3}\,e^{2\sqrt{3}\,\Phi}F_{\alpha\beta}F^{\alpha\beta}=0.
	\ee
	
	\subsection{Uplift and boost}
	
	Starting from a stationary axisymmetric Ricci-flat seed
	$\tilde g_{\mu\nu}$, we uplift it to five dimensions by adjoining a flat
	compact direction $z$:
	\be \label{eq:ds5}
	ds_5^2 = \tilde g_{\mu\nu}\,dx^\mu dx^\nu + dz^2
	= G_{MN}\,dx^M dx^N,
	\ee
	with four-dimensional coordinates
	\[
	x^\nu=\{t,x,y,\phi\},
	\]
	and five-dimensional ones $x^N=\{x^\mu,z\}$.
	
	A Lorentz boost along the compact direction,
	\be
	\begin{aligned}\label{eq:boost}
	dt &\rightarrow \cosh\alpha\,dt + \sinh\alpha\,dz,\\[1mm]
	dz &\rightarrow \cosh\alpha\,dz + \sinh\alpha\,dt,
	\end{aligned}
	\ee
	introduces momentum along $z$. After reduction, the boosted five-dimensional
	metric can be cast in the KK form
	\be \label{eq:boosted5d}
	ds_5^2 = H^{-1}\,g_{\mu\nu}\,dx^\mu dx^\nu
	+H^2\bigl(dz+2A_\mu dx^\mu\bigr)^2,
	\ee
	with
	\be \label{eq:Hdef}
	H^2=\cosh^2\!\alpha+\tilde g_{tt}\sinh^2\!\alpha.
	\ee
	
	\subsection{Reduced four-dimensional fields}
	
	The resulting Einstein-frame metric may be written in orthonormal form as
	\be
	ds^2 = \eta_{(i)(j)}\,\boldsymbol{\omega}^{(i)}\,\boldsymbol{\omega}^{(j)},
	\ee
	where
	\be \label{eq:tetrads}
	\begin{aligned}
	\boldsymbol{\omega}^{(0)} &= \sqrt{-\frac{\tilde g_{tt}}{H}}
	\left(dt+\cosh\alpha\,\frac{\tilde g_{t\phi}}{\tilde g_{tt}}\,d\phi\right),\\[2mm]
	\boldsymbol{\omega}^{(1)} &= \sqrt{H\,\tilde g_{xx}}\,dx,\\[2mm]
	\boldsymbol{\omega}^{(2)} &= \sqrt{H\,\tilde g_{yy}}\,dy,\\[2mm]
	\boldsymbol{\omega}^{(3)} &= \sqrt{\frac{H}{\tilde g_{tt}}
	\left(\tilde g_{\phi\phi}\tilde g_{tt}-\tilde g_{t\phi}^{\,2}\right)}\,d\phi.
	\end{aligned}
	\ee
	The accompanying matter fields are
	\be \label{eq:gauge}
	A_\mu dx^\mu=
	\frac{\sinh\alpha\Bigl[(1+\tilde g_{tt})\cosh\alpha\,dt
	+\tilde g_{t\phi}\,d\phi\Bigr]}{2H^2},
	\ee
	and
	\be \label{eq:dilaton}
	\Phi=\frac{\sqrt{3}}{2}\ln H.
	\ee
	By construction, the fields
	$\{g_{\mu\nu},A_\mu,\Phi\}$ obtained in this way solve
	Eqs.~\eqref{eq:Rmn}--\eqref{eq:Phi} whenever the seed metric is
	Ricci-flat; for the TS$_2$ seed below we have verified this by direct
	symbolic computation of all components of
	\eqref{eq:Rmn}--\eqref{eq:Phi}, in \textsc{Maple} and independently in
	\textsc{SymPy} (the \textsc{SymPy} code was written with the assistance of
	an AI model; see the declaration before the references).\footnote{Some authors write the coupling as
	$e^{-2\sqrt3\Phi}F^2$; that form is related to ours by the field
	redefinition $\Phi\to-\Phi$, which flips the sign of the dilaton charge but
	changes no physical statement below. We follow the sign conventions of
	Ref.~\cite{Aliev:2008wv}, under which $\Sigma>0$ for the boosted solutions.}
	
\section{The Tomimatsu--Sato $\delta=2$ Vacuum Solution}
\label{sec:TS}

The TS family~\cite{Tomimatsu:1972zz,Tomimatsu:1973gq} constitutes a class of
exact stationary axisymmetric vacuum solutions of Einstein's equations. The
discrete parameter $\delta$ labels the members of the family: $\delta=1$
recovers the Kerr spacetime, while $\delta=2$ gives the simplest non-Kerr
member.

\subsection{Prolate spheroidal coordinates}

The TS metrics are naturally written in prolate spheroidal coordinates
$(x,y)$ with ranges
\be
x\geq 1, \qquad -1\leq y\leq 1.
\ee
These coordinates are related to Weyl--Papapetrou coordinates by
\be \label{eq:prolate}
\rho=\kappa\sqrt{(x^2-1)(1-y^2)}, \qquad z=\kappa xy,
\ee
where $\kappa$ is the length scale of the solution.

In the Ernst--Perj\'es representation, the vacuum TS$_2$ metric takes the form
\be \label{eq:TS2metric}
ds^2 = \kappa^2 f^{-1}
\left[
e^{2\gamma}(x^2-y^2)
\left(\frac{dx^2}{x^2-1}+\frac{dy^2}{1-y^2}\right)
+(x^2-1)(1-y^2)\,d\phi^2
\right]
-f\,(dt-\omega d\phi)^2.
\ee
The metric functions are
\bea
f &=& \frac{\mu^2-(x^2-1)(1-y^2)\sigma^2}
{\mu^2+\mu\nu-(1-y^2)\bigl[(x^2-1)\sigma^2-\sigma\tau\bigr]},
\label{eq:TS2f}\\[2mm]
e^{2\gamma} &=& \frac{\mu^2-(x^2-1)(1-y^2)\sigma^2}{p^4(x^2-y^2)^4},
\label{eq:TS2gamma}\\[2mm]
\omega &=& -\kappa(1-y^2)
\frac{(x^2-1)\sigma\nu+\mu\tau}{\mu^2-(x^2-1)(1-y^2)\sigma^2},
\label{eq:TS2omega}
\eea
with auxiliary polynomials
\bea
\mu &=& p^2(x^2-1)^2+q^2(1-y^2)^2,
\label{eq:TS2mu}\\[1mm]
\sigma &=& 2pq(x^2-y^2),
\label{eq:TS2sigma}\\[1mm]
\nu &=& 4x(px^2+2x+p),
\label{eq:TS2nu}\\[1mm]
\tau &=& -4q\,p^{-1}(1-y^2)(px+1),
\label{eq:TS2tau}
\eea
where
\be \label{eq:TS2pq}
p^2+q^2=1.
\ee

For comparison with the canonical Weyl--Papapetrou form,
\be \label{eq:WP}
ds^2=-f(dt-\omega d\phi)^2+f^{-1}\left[e^{2\gamma}(d\rho^2+dz^2)+\rho^2d\phi^2\right],
\ee
the TS family may also be written in terms of three polynomials $A$, $B$, and
$C$ through
\be \label{eq:ABCform}
f=\frac{A}{B}, \qquad
\omega=\frac{4\kappa q}{p}\,\frac{(1-y^2)\,C}{A}, \qquad
e^{2\gamma}=\frac{A}{p^{4}(x^2-y^2)^{4}},
\ee
where the polynomials are
\bea
A &=& p^4(x^2-1)^4+q^4(1-y^2)^4
-2p^2q^2(x^2-1)(1-y^2)
\nonumber\\
&&\times\Bigl\{2(x^2-1)^2+2(1-y^2)^2+3(x^2-1)(1-y^2)\Bigr\},
\\[2mm]
B &=& \Bigl\{p^2(x^2+1)(x^2-1)-q^2(y^2+1)(1-y^2)+2px(x^2-1)\Bigr\}^2
\nonumber\\
&&\qquad +4q^2y^2\Bigl\{px(x^2-1)+(px+1)(1-y^2)\Bigr\}^2,
\\[2mm]
C &=& -p^3x(x^2-1)\Bigl\{2(x^2+1)(x^2-1)+(x^2+3)(1-y^2)\Bigr\}
\nonumber\\
&&\qquad -p^2(x^2-1)\Bigl\{4x^2(x^2-1)+(3x^2+1)(1-y^2)\Bigr\}
+q^2(px+1)(1-y^2)^3.
\eea

\begin{figure}[tp]
\centering
\includegraphics[width=\textwidth]{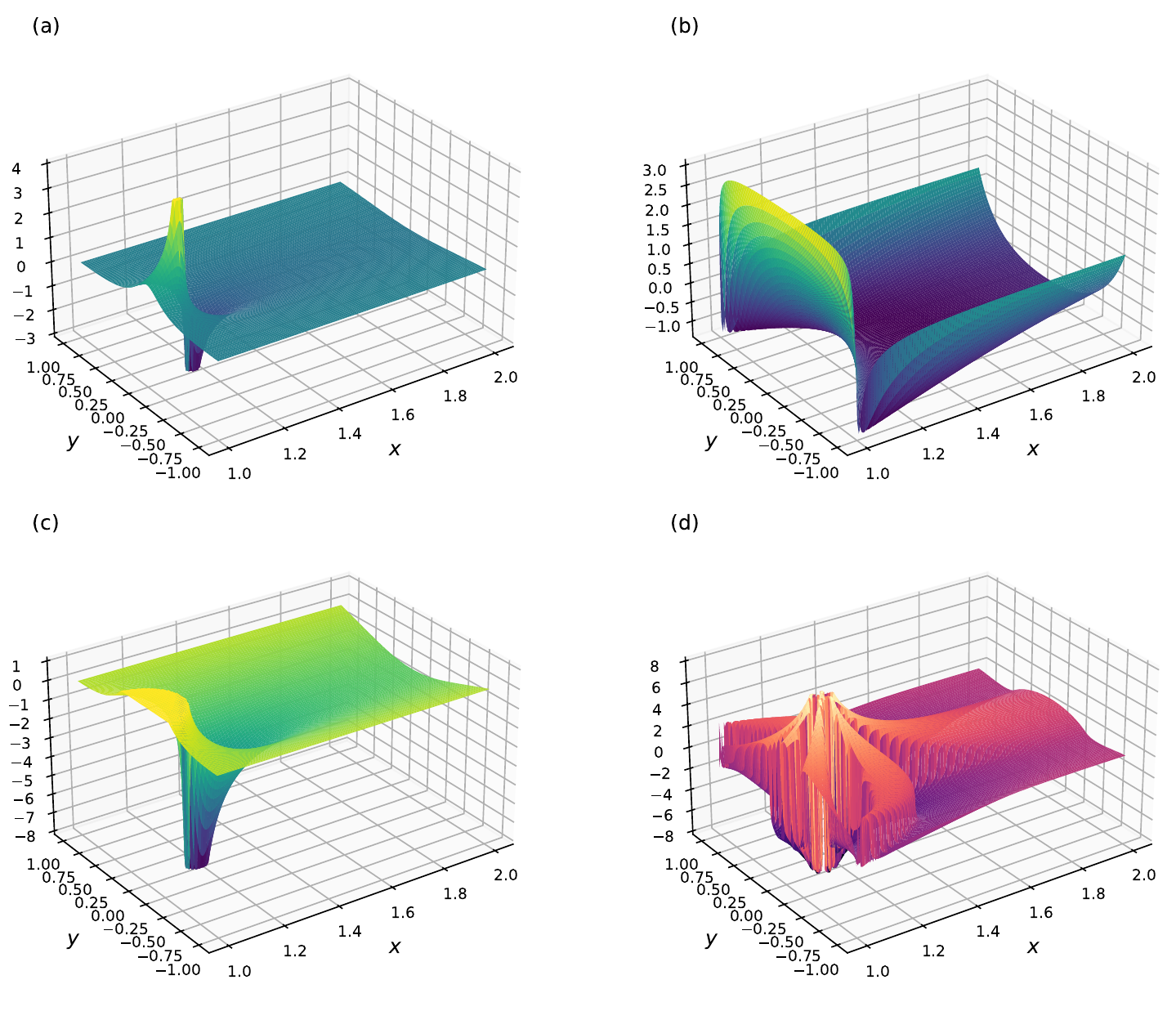}
\caption{Metric functions of the vacuum TS$_2$ seed for $p=0.6$, $\kappa=0.3$
($m=1$), plotted over $1\le x\le2$, $|y|\le1$; the vertical axis of each panel
is the function named below. (a) $f$, clipped to $[-3,4]$, and (b)
$e^{2\gamma}$, clipped to $[-2,4]$: both diverge at the ring singularity; note
the ergoregion $f<0$ and the lens $f>1$ near the equator. (c) The
frame-dragging component $\tilde g_{t\phi}=f\omega$, clipped to $[-8,1]$. (d)
The Kretschmann scalar on the signed logarithmic scale
$\mathrm{sgn}(K)\log_{10}(1+|K|)$, clipped to $[-8,8]$; the vacuum Ricci
scalar vanishes identically and is not shown.}
\label{fig:seed}
\end{figure}

Figure~\ref{fig:seed} displays the seed metric functions and the Kretschmann
invariant for the parameters used throughout the paper, $p=0.6$ and
$\kappa=0.3$ (seed mass $m=1$). Several features of the global structure
discussed in Section~\ref{sec:properties} are already visible here. The
function $f$ is negative over a large ergoregion and exceeds unity in a thin
lens between the singular rod $x=1$, $|y|<1$ (whose endpoints are the two
degenerate horizons, Sec.~\ref{subsec:horizon}) and the equatorial ring
singularity; it is this $f>1$ region that the boost will turn into a
curvature-singular boundary. The
conformal factor $e^{2\gamma}$ shares the numerator $A$ with $f$ and therefore
changes sign together with it, so that $\gamma$ itself is complex inside the
ergoregion; we plot $e^{2\gamma}$ for that reason. The frame-dragging
component $\tilde g_{t\phi}=f\omega$ is regular everywhere except at the ring,
where the denominator $B$ of \eqref{eq:ABCform} vanishes (the function
$\omega$ alone diverges on the ergosurface $f=0$ and is less informative).
Finally, since the seed is Ricci-flat all its Ricci invariants vanish, and we
use the Kretschmann scalar $K=R_{abcd}R^{abcd}$ to visualize the curvature;
it is shown on a signed logarithmic
scale: it is smooth over the whole domain, diverges only at the ring
singularity, and approaches $48m^2/r^6$ at large distance with $r=\kappa x+m$.

\subsection{Asymptotic charges and multipole moments of the seed}
\label{subsec:seedcharges}

Expanding the metric functions~\eqref{eq:TS2f} and \eqref{eq:TS2omega} at
large $x$ gives
\be \label{eq:seedasym}
f = 1-\frac{4}{p\,x}+\mathcal{O}(x^{-2}), \qquad
f\omega = -\frac{8\kappa q\,(1-y^2)}{p^2\,x}+\mathcal{O}(x^{-2}),
\ee
so that, with the asymptotic radial coordinate $r\simeq\kappa x$ and
$y=\cos\theta$, comparison with
$\tilde g_{tt}\simeq -1+2m/r$ and $\tilde g_{t\phi}\simeq -2J_0\sin^2\!\theta/r$
identifies the seed mass and angular momentum,
\be \label{eq:seedMJ}
m=\frac{2\kappa}{p}, \qquad J_0=m^2 q .
\ee
These are the familiar TS relations: the dimensionless spin of the seed is
$j_0\equiv J_0/m^2=q$, and $\kappa=mp/2$ plays the role of
$\sqrt{m^2-a^2}/\,$(horizon-scale) parameter, with $p\to1$ the static limit
and $p\to0$ the extreme-rotation limit.

The higher moments are conveniently obtained from the Ernst potential
\cite{Ernst:1967wx}. For TS$_2$,
\be \label{eq:xiTS2}
\xi=\frac{p^2x^4+q^2y^4-1-2ipqxy\,(x^2-y^2)}
{2px(x^2-1)-2iqy(1-y^2)},
\ee
in terms of which $f=(\xi\bar\xi-1)/[(\xi+1)(\bar\xi+1)]$; we have verified
that \eqref{eq:xiTS2} reproduces \eqref{eq:TS2f} identically. In this convention
$\xi\to\infty$ at spatial infinity; the potential that decays there, and to which
the Fodor--Hoenselaers--Perj\'es algorithm applies, is its reciprocal
$\tilde\xi=1/\xi=(1-\mathcal E)/(1+\mathcal E)$, with $\mathcal E$ the Ernst potential. On the symmetry
axis $y=1$ the potential collapses to the rational function
\be \label{eq:xiaxis}
\xi(x,y{=}1)=\frac{p\,(x^2+1)-2iqx}{2x},
\ee
and the Geroch--Hansen moments \cite{Geroch:1970cd,Hansen:1974zz} follow from
the Fodor--Hoenselaers--Perj\'es algorithm \cite{Fodor:1989} applied to
$\tilde\xi=1/\xi$: writing $\zeta=\kappa x$ for the Weyl coordinate along the
axis, one expands
\be \label{eq:FHP}
\frac{1}{\xi(\zeta)}=\sum_{j\geq0}\frac{m_j}{\zeta^{\,j+1}},
\qquad
m_0=M_0,\quad m_1=iS_1,\quad m_2=M_2,\quad m_3=iS_3,
\ee
the low-order moments being uncorrected. Carrying out the expansion of
\eqref{eq:xiaxis} gives
\be \label{eq:seedmoments}
M_0=m,\qquad
S_1=J_0=m^2q,\qquad
M_2=-m^3\Bigl(q^2+\frac{p^2}{4}\Bigr),\qquad
S_3=-m^4q\Bigl(q^2+\frac{p^2}{2}\Bigr).
\ee
These values are not new. The mass and angular momentum are those of
\cite{Tomimatsu:1972zz,Tomimatsu:1973gq}; the quadrupole is the $\delta=2$
case of the general-$\delta$ formula of \cite{Tomimatsu:1973gq},
$M_2=-m^3[q^2+(\delta^2-1)p^2/(3\delta^2)]$, with the sign as corrected by
Manko \cite{Manko:2012ts}, who recomputed it with the same algorithm and
wrote it as $-\tfrac14m^3(1+3q^2)$; and the current octupole is the vacuum
limit of the moments of the charged, magnetized TS$_2$ solution
\cite{Manko:1999ts}, whose $M_3=-8ima(m^2+a^2)$ in the normalization of that
paper (total mass $2m$, $J=4ma$) is $S_3=-Ma(M^2+a^2)/2$ in ours, i.e.\
\eqref{eq:seedmoments}. We quote them here because the charged solution is
compared with them below.
For Kerr with the same $(m,q)$ one has $M_2^{\rm Kerr}=-m^3q^2$ and
$S_3^{\rm Kerr}=-m^4q^3$, so the $\delta=2$ deformation is encoded in the
excesses
\be \label{eq:seedexcess}
\Delta M_2=-\frac{p^2}{4}\,m^3,
\qquad
\Delta S_3=-\frac{p^2}{2}\,m^4 q .
\ee
The quadrupole excess $\Delta M_2$ is the leading gauge-invariant
distinction between TS$_2$ and Kerr, and it is this quantity whose fate under
the KK boost we track in Section~\ref{subsec:charges}.

	\section{Charged Tomimatsu--Sato Solution in Kaluza--Klein Theory}
	\label{sec:charged_TS}
	
	We now apply the boost construction of Section~\ref{sec:KK} to the TS$_2$ seed.
	The reduced fields
	\[
	(g_{\mu\nu},A_\mu,\Phi)
	\]
	provide an exact charged solution of the four-dimensional KK equations.
	
	\subsection{Construction}
	
	For the seed metric~\eqref{eq:TS2metric}, the nonvanishing coordinate-basis
	components are
	\bea
	\tilde g_{tt} &=& -f, \\
	\tilde g_{t\phi} &=& f\omega, \\
	\tilde g_{\phi\phi} &=& \kappa^2 f^{-1}(x^2-1)(1-y^2)-f\omega^2, \\
	\tilde g_{xx} &=& \kappa^2 f^{-1} e^{2\gamma}\frac{x^2-y^2}{x^2-1}, \\
	\tilde g_{yy} &=& \kappa^2 f^{-1} e^{2\gamma}\frac{x^2-y^2}{1-y^2}.
	\eea
	Hence the boost factor is
	\be \label{eq:HTS}
	H^2 = \cosh^2\!\alpha-f\sinh^2\!\alpha
	= 1 + (1-f)\sinh^2\!\alpha.
	\ee
	
	The tetrads in Eq.~\eqref{eq:tetrads} become
	\bea
	\boldsymbol{\omega}^{(0)} &=& \sqrt{\frac{f}{H}}\,(dt-\cosh\alpha\,\omega\,d\phi), \\
	\boldsymbol{\omega}^{(1)} &=& \kappa\sqrt{\frac{H}{f}e^{2\gamma}\frac{x^2-y^2}{x^2-1}}\;dx, \\
	\boldsymbol{\omega}^{(2)} &=& \kappa\sqrt{\frac{H}{f}e^{2\gamma}\frac{x^2-y^2}{1-y^2}}\;dy, \\
	\boldsymbol{\omega}^{(3)} &=& \kappa\sqrt{\frac{H}{f}(x^2-1)(1-y^2)}\;d\phi.
	\eea
	Accordingly, the Einstein-frame metric reads
	\be \label{eq:KKTSmetric}
	ds^2 = -\frac{f}{H}(dt-\cosh\alpha\,\omega\,d\phi)^2
	+\kappa^2\frac{H}{f}
	\left[
	e^{2\gamma}(x^2-y^2)
	\left(\frac{dx^2}{x^2-1}+\frac{dy^2}{1-y^2}\right)
	+(x^2-1)(1-y^2)\,d\phi^2
	\right].
	\ee
	The Maxwell potential is
	\be \label{eq:gaugeTS}
	A = A_t\,dt + A_\phi\,d\phi,
	\ee
	with
	\bea
	A_t &=& \frac{\cosh\alpha\sinh\alpha\,(1-f)}{2H^2},
	\label{eq:At}\\
	A_\phi &=& \frac{\sinh\alpha\,f\omega}{2H^2}.
	\label{eq:Aphi}
	\eea
	The dilaton is
	\be \label{eq:dilatonTS}
	\Phi = \frac{\sqrt{3}}{2}\ln H.
	\ee
	When $\alpha=0$, one has $H=1$, $A_\mu=0$, and $\Phi=0$, so the solution
	reduces smoothly to the vacuum TS$_2$ geometry.
	
\begin{figure}[tp]
\centering
\includegraphics[width=\textwidth]{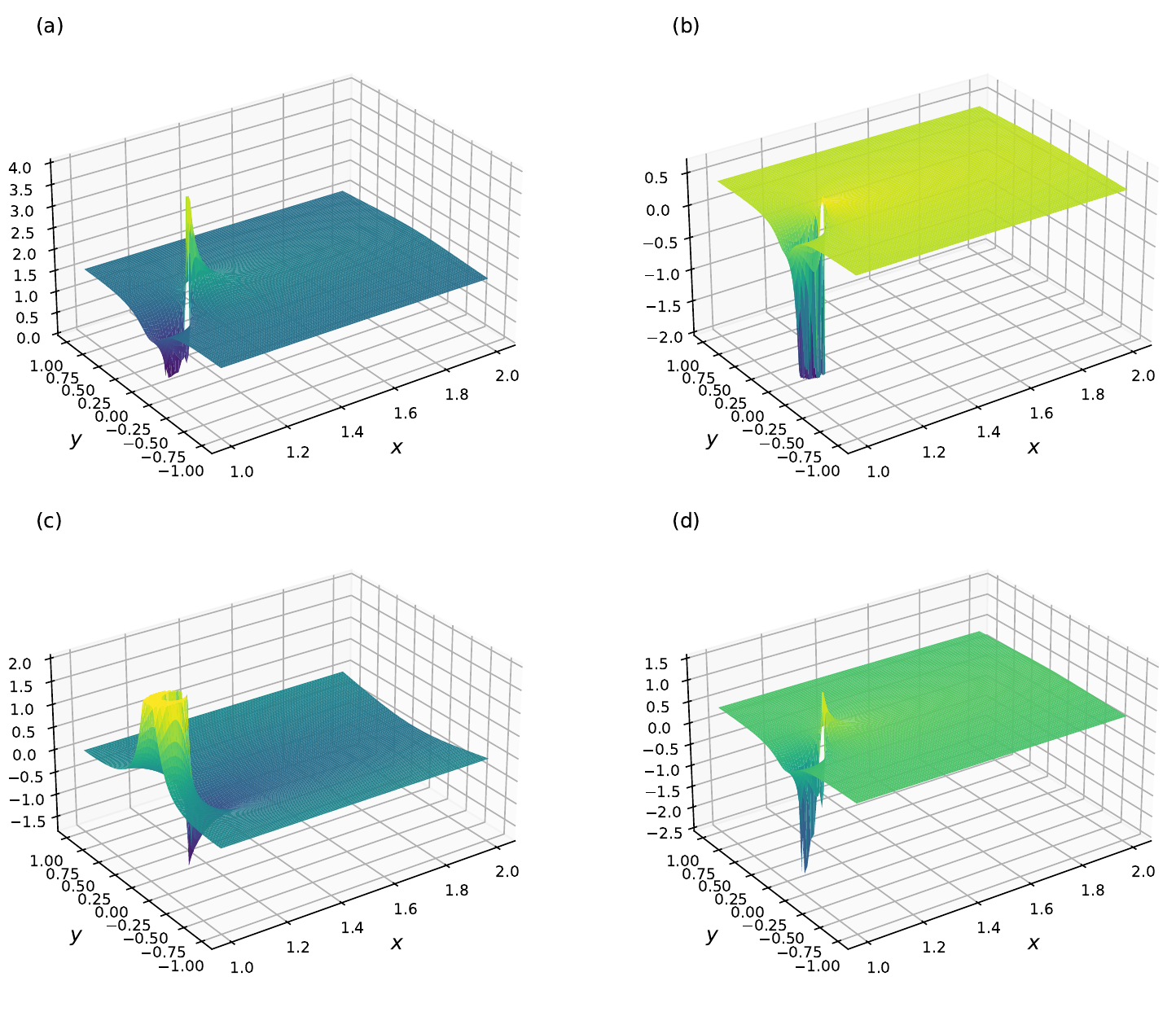}
\caption{Fields of the charged solution for $p=0.6$, $\kappa=0.3$,
$\alpha=1$, plotted over $1\le x\le2$, $|y|\le1$ on the physical domain
\eqref{eq:domain}; the vertical axis of each panel is the field named below.
(a) The boost function $H$, clipped to $[0,4]$. (b) The Maxwell potential
$A_t$, clipped to $[-2,4]$. (c) The Maxwell potential $A_\phi$, clipped to
$[-2,2]$. (d) The dilaton $\Phi=\tfrac{\sqrt3}{2}\ln H$, clipped to
$[-2.5,1.5]$. The hole in each surface is the excised lens $H^2\le0$
(Section~\ref{subsec:global}); the potentials and the dilaton diverge on its
boundary.}
\label{fig:charged}
\end{figure}

	The boost function $H$ and the matter fields are shown in
	Fig.~\ref{fig:charged} for $\alpha=1$. Two features deserve comment. First,
	all four surfaces have a hole near $x\simeq1.1$, $y\simeq0$: this is the
	region $H^2\le0$, where $f>\coth^2\!\alpha$ and the fields cease to be real;
	it is not a plotting artifact but the excised lens bounded by the new
	curvature singularity discussed in Section~\ref{subsec:global}, and the
	surfaces are deliberately masked to the physical domain
	\eqref{eq:domain}. Second, away from that lens the fields are smooth: $H$
	tends to unity at infinity and exceeds $\cosh\alpha$ throughout the
	ergoregion, $A_t$ and $A_\phi$ fall off as $Q/r$ and as a magnetic dipole,
	and $\Phi=\tfrac{\sqrt3}{2}\ln H$ is positive in the ergoregion, negative
	near the lens, and diverges to $-\infty$ on its boundary.
	
	We note for later use that
	the charged metric \eqref{eq:KKTSmetric} is again of Weyl--Papapetrou
	form~\eqref{eq:WP} in the \emph{same} coordinates $(\rho,z)$ of
	\eqref{eq:prolate}, with
	\be \label{eq:WPcharged}
	f'=\frac{f}{H},\qquad
	\omega'=\cosh\alpha\,\omega,\qquad
	\gamma'=\gamma .
	\ee
	
	\subsection{Static limit: charged dilatonic Zipoy--Voorhees}
	\label{subsec:static}
	
	It is instructive to record the non-rotating limit of the seed, obtained by
	sending $q\to0$ (equivalently $p\to1$, so that $J_0=m^2q\to0$). In this limit
	the auxiliary polynomials \eqref{eq:TS2sigma} and \eqref{eq:TS2tau} vanish,
	$\sigma,\tau\to0$, and the seed functions
	\eqref{eq:TS2f}--\eqref{eq:TS2omega} collapse to
	\be \label{eq:ZVlimit}
	f\;\longrightarrow\;\left(\frac{x-1}{x+1}\right)^{2},
	\qquad
	e^{2\gamma}\;\longrightarrow\;\left(\frac{x^2-1}{x^2-y^2}\right)^{4},
	\qquad
	\omega\;\longrightarrow\;0 ,
	\ee
	which is exactly the static Zipoy--Voorhees (Weyl $\gamma$-)metric of
	distortion parameter $\delta=2$: the potential exponent is $\delta$ and the
	conformal exponent is $\delta^2=4$ \cite{Zipoy:1966btu,Voorhees:1970ywo}.
	Applying the boost \eqref{eq:HTS}--\eqref{eq:dilatonTS} to \eqref{eq:ZVlimit},
	with $\omega=0$ so that $A_\phi=0$, gives a \emph{static} charged dilatonic
	solution: a purely electric potential $A_t=\tfrac12\cosh\alpha\sinh\alpha\,
	(1-f)/H^2$, dilaton $\Phi=\tfrac{\sqrt3}{2}\ln H$, and Einstein-frame metric
	\eqref{eq:KKTSmetric} with $\omega=0$. This is precisely the charged
	dilatonic Zipoy--Voorhees $\delta=2$ metric constructed and analyzed
	(exact form, equatorial geodesics, circular orbits, and Melvin immersion) in
	\cite{Siahaan:2026zvd,Siahaan:2026mzv}, specialized to the Kaluza--Klein
	coupling $a=\sqrt3$. The solution presented here is therefore the rotating
	generalization of that family: the seed parameter $q$ restores angular
	momentum, activates the magnetic sector $A_\phi\neq0$, and --- as we shall
	see --- deforms the multipole and causal structure and, in particular,
	generates the region $f>1$ responsible for the reduction singularity of
	Sec.~\ref{subsec:global}, which is absent in the static limit.
	
	\section{Geometric Properties}
	\label{sec:properties}
	
	\subsection{Asymptotic behavior, conserved charges, and multipole moments}
	\label{subsec:charges}
	
\subsubsection*{Large-distance expansion}

The conserved charges are read off from the behaviour of the fields at
spatial infinity, $x\to\infty$. We use the radial coordinate $r\equiv\kappa x$;
any shift $r\to r+{\rm const}$ changes nothing at the order that defines the
charges, and the Boyer--Lindquist-like coordinate $r+m$ would make the seed
potential Schwarzschild-like to the next order, $f=1-2m/(r+m)+\mathcal O(r^{-3})$.
Expanding \eqref{eq:TS2f}--\eqref{eq:TS2omega} to second order (the first
order is \eqref{eq:seedasym}),
\be \label{eq:seedasym2}
f=1-\frac{2m}{r}+\frac{2m^2}{r^2}+\mathcal{O}(r^{-3}),\qquad
f\omega=-\frac{2J_0\sin^2\!\theta}{r}\Bigl(1-\frac{m}{r}\Bigr)+\mathcal{O}(r^{-3}),
\ee
with $y=\cos\theta$, $m=2\kappa/p$ and $J_0=m^2q$. Both $1/r^2$ coefficients
are angle-independent apart from the $\sin^2\!\theta$ of $g_{t\phi}$: this is
the statement that, at this order, the prolate chart already satisfies Thorne's
asymptotically Cartesian condition for $g_{tt}$ (no harmonic beyond $\ell=0$ at
order $1/r^2$; reflection symmetry removes the mass dipole), and it is what
makes the quadrupole extraction below tractable. Inserting \eqref{eq:seedasym2} into the boost formulae
\eqref{eq:HTS} and \eqref{eq:KKTSmetric}--\eqref{eq:dilatonTS} gives, with
$M$, $J$, $Q$, $\Sigma$ and $\mu_{\rm mag}$ as defined in
\eqref{eq:Mcharge}--\eqref{eq:mudipole} below,
\be \label{eq:asymfields}
\begin{aligned}
H&=1+\frac{m\sinh^2\!\alpha}{r}-\frac{mM\sinh^2\!\alpha}{r^2}+\mathcal{O}(r^{-3}),\\
-g_{tt}&=1-\frac{2M}{r}+\frac{2M^2+Q^2}{r^2}+\mathcal{O}(r^{-3}),\\
g_{t\phi}&=-\frac{2J\sin^2\!\theta}{r}\Bigl(1-\frac{m\cosh^2\!\alpha}{r}\Bigr)+\mathcal{O}(r^{-3}),\\
A_t&=\frac{Q}{r}\Bigl(1-\frac{m\cosh2\alpha}{r}\Bigr)+\mathcal{O}(r^{-3}),\qquad
A_\phi=-\frac{\mu_{\rm mag}\sin^2\!\theta}{r}\Bigl(1-\frac{m\cosh2\alpha}{r}\Bigr)+\mathcal{O}(r^{-3}),\\
\Phi&=\frac{\Sigma}{r}\Bigl(1-\frac{m\cosh^2\!\alpha}{r}\Bigr)+\mathcal{O}(r^{-3}).
\end{aligned}
\ee
We verified \eqref{eq:asymfields} symbolically in $p$ and $\alpha$. The
charges are defined by the leading terms in the standard way: $M$ by
$-g_{tt}\to1-2M/r$ (equivalently by the Komar integral of $\partial_t$ at
infinity), $J$ by $g_{t\phi}\to-2J\sin^2\!\theta/r$ (the Komar integral of
$\partial_\phi$), $Q$ by $A_t\to Q/r$ (Gauss' law,
$Q=-\frac1{4\pi}\oint_\infty e^{2\sqrt3\Phi}\star F$), $\mu_{\rm mag}$ by
$A_\phi\to-\mu_{\rm mag}\sin^2\!\theta/r$, and the dilaton charge by
$\Phi\to\Sigma/r$. Note that, unlike in vacuum, the Komar integral of
$\partial_t$ over a sphere of finite radius is not equal to $M$: the
Maxwell--dilaton stress outside the sphere contributes, and the integral
approaches $M$ only as $r\to\infty$. The electric charge, by contrast, is
exactly conserved.

\subsubsection*{The conserved charges}

Reading off the $1/r$ coefficients of \eqref{eq:asymfields} yields
\bea
M &=& \frac{m}{2}\left(1+\cosh^2\!\alpha\right),
\label{eq:Mcharge}\\
J &=& m^2 q\,\cosh\alpha ,
\label{eq:Jcharge}\\
Q &=& m\,\sinh\alpha\cosh\alpha ,
\label{eq:Qcharge}\\
\Sigma &=& \frac{\sqrt{3}}{2}\,m\,\sinh^2\!\alpha ,
\label{eq:Scharge}
\eea
where $Q$ is normalized by $A_t\to Q/r$ and the dilaton charge by
$\Phi\to\Sigma/r$.\footnote{\label{fn:signs}Our gauge potential
\eqref{eq:gauge} has the opposite overall sign to the Kerr--Newman convention
$A_t\to-Q/r$. Correspondingly $A_\phi\to-\mu_{\rm mag}\sin^2\!\theta/r$, the
charge enclosed by a closed surface $S$ with the outward orientation is
$Q(S)=-\frac{1}{4\pi}\oint_Se^{2\sqrt3\Phi}\star F$, and the signs of $Q$,
$\mu_{\rm mag}$, $g$ and $\Phi_e$ below are all consistent with this choice.}
The magnetic potential \eqref{eq:Aphi} carries a magnetic dipole moment
\be \label{eq:mudipole}
\mu_{\rm mag}=m^2 q\,\sinh\alpha = J_0\sinh\alpha .
\ee
Each of these follows from a single line of \eqref{eq:asymfields}. The mass
receives two contributions, $M=m+\tfrac12m\sinh^2\!\alpha$: the seed mass,
which survives in $-g_{tt}=-f/H$ through the factor $f$, and half of the
$1/r$ coefficient of $H$, which is the gravitational mass of the Maxwell and
dilaton fields generated by the boost; the same coefficient of $H$ gives the
dilaton charge, $\Sigma=\tfrac{\sqrt3}{2}m\sinh^2\!\alpha$, so that
\be \label{eq:mfromMS}
m=M-\frac{\Sigma}{\sqrt3}
\ee
expresses the seed mass through the asymptotic charges. The angular momentum
and the magnetic dipole are both proportional to the seed angular momentum
$J_0$, with the factors $\cosh\alpha$ and $\sinh\alpha$ of the boost, since
$g_{t\phi}$ and $A_\phi$ are the $dt$- and $dz$-components of the single
five-dimensional cross term $\tilde g_{t\phi}\,d\phi\,(\cosh\alpha\,dt
+\sinh\alpha\,dz)$; and the electric charge is the Kaluza--Klein momentum
generated by the boost, which vanishes with $\alpha$. In five dimensions all
four charges are therefore the mass $m$ and the angular momentum $J_0$ of
the seed seen from a boosted frame, which is why they depend on the seed
only through $(m,J_0)$ and not on $\delta$.

\subsubsection*{Relations among the charges}

The four charges are built from the three numbers $m$, $q$ and $\alpha$, so
they satisfy one relation. It is the KK constraint, in the form given in
\cite{Aliev:2008wv},
\be \label{eq:KKconstraint}
\Sigma^2+\sqrt{3}\,M\,\Sigma=\frac{3}{2}\,Q^2,\qquad\text{i.e.}\qquad Q^2=\frac{2\Sigma}{\sqrt{3}}\left(M+\frac{\Sigma}{\sqrt{3}}\right),
\ee
which is verified by direct substitution of
\eqref{eq:Mcharge}--\eqref{eq:Scharge}. Solving it for the dilaton charge,
\be \label{eq:Sigmasolved}
\Sigma=\frac{\sqrt3}{2}\Bigl(\sqrt{M^2+2Q^2}-M\Bigr)=\frac{\sqrt3}{2}\,Q\tanh\alpha,
\qquad
\tanh\alpha=\frac{\sqrt{M^2+2Q^2}-M}{Q},
\ee
shows that $\Sigma$ is secondary hair, fixed by $M$ and $Q$, and that the
boost parameter itself is a function of the asymptotic charges. The
gyromagnetic ratio is
\be \label{eq:gfactor}
g=\frac{2M\mu_{\rm mag}}{QJ}=2-\tanh^2\!\alpha
=2-\frac{\bigl(\sqrt{M^2+2Q^2}-M\bigr)^2}{Q^2},
\ee
which decreases monotonically from the Dirac value $g=2$ in the limit of
vanishing charge to $g=1$ for $\alpha\to\infty$; it is independent of $q$,
i.e.\ of how fast the seed rotates. The remaining dimensionless ratios are
also simple functions of $\alpha$ alone,
\be \label{eq:ratios}
\frac{Q}{M}=\frac{\sinh2\alpha}{1+\cosh^2\!\alpha},\qquad
\frac{\Sigma}{M}=\frac{\sqrt3\sinh^2\!\alpha}{1+\cosh^2\!\alpha},\qquad
\frac{J}{M^2}=q\,\frac{4\cosh\alpha}{(1+\cosh^2\!\alpha)^2}.
\ee
The first increases monotonically from $0$ to $2$, and $2$ is precisely the
extremal value $Q^2=(1+a^2)M^2$ of the static dilaton black holes with
coupling $a=\sqrt3$ \cite{Gibbons:1987ps,Garfinkle:1990qj}: the boost
approaches the extremal bound but reaches it only as $\alpha\to\infty$ (at
fixed $Q$ this is the limit $m\to0$). In the same limit $\Sigma/M\to\sqrt3$.
The last ratio, the dimensionless spin, is $q$ in the vacuum limit and
decreases monotonically to zero: boosting adds mass faster than angular
momentum, so the charged solution is always further from the Kerr bound
than its seed. For the parameters used in the figures, $p=3/5$ and $\alpha=1$,
the charges are
$M=1.6905\,m$, $J=1.2345\,m^2$, $Q=1.8134\,m$, $\Sigma=1.1961\,m$,
$\mu_{\rm mag}=0.9402\,m^2$, $g=1.420$, with $Q/M=1.073$ and $J/M^2=0.432$
against $q=0.8$ for the seed.

Equations \eqref{eq:Mcharge}--\eqref{eq:ratios} coincide with the
corresponding relations for the boosted Kerr (KK) black hole
\cite{Frolov:1987rj,Rasheed:1995zv,Larsen:1999pp,Aliev:2008wv,Knoll:2015nva}
with the same seed mass and angular momentum (there $J_0=ma$), and, by the
argument given above, with those of the KK dressing of any stationary
axisymmetric vacuum seed with the same $(m,J_0)$: at this order the boost
probes only the universal $1/r$ tail of the seed, and the $\delta=2$
structure is invisible. The $1/r^2$ terms of \eqref{eq:asymfields} are
likewise universal, being fixed by $(m,J_0,\alpha)$ and free of angular
harmonics beyond the $\sin^2\!\theta$ of the rotational components. The
imprint of the seed first appears at quadrupole order, i.e.\ in the
$P_2(\cos\theta)/r^3$ term of $g_{tt}$, to which we now turn.

\subsubsection*{Mass quadrupole and its boost-scaling law}

For the charged solution the vacuum Ernst machinery of
Section~\ref{subsec:seedcharges} is no longer directly available: the twist
one-form of the timelike Killing vector fails to be closed in the presence of
the Maxwell--dilaton stress, and a naive application of the vacuum
Fodor--Hoenselaers--Perj\'es reading to the pair $(f',\omega')$ of
\eqref{eq:WPcharged} produces spurious parity-violating moments at
$\mathcal{O}(Q^2)$. The appropriate invariant notion is instead provided by
Mayerson's generalization of the Geroch--Hansen formalism to stationary
non-vacuum spacetimes \cite{Mayerson:2022ekj}, which shows that, for a
spacetime that is sufficiently smooth at spatial infinity and admits a
well-defined improved twist potential, the moments read off in asymptotically
Cartesian mass-centred (ACMC) coordinates in the sense of Thorne
\cite{Thorne:1980ru} are the invariant (generalized Geroch--Hansen)
multipoles. The topological part of these assumptions --- exactness of a
two-form built from the stress tensor --- can be proved for the present
solution, and the smoothness part holds in the same sense as for Kerr
(Appendix~\ref{app:acmc}).
We therefore construct an ACMC chart for the Einstein-frame metric: the
radial and angular coordinates are fixed, order by order in $1/R$, by the
requirement that the \emph{Cartesian} components contain no harmonic of order
$\ell\geq1$ at $\mathcal{O}(1/R)$ and none of order $\ell\geq2$ at
$\mathcal{O}(1/R^2)$ --- for the spatial metric we impose the stronger,
sufficient condition of isotropy, $g_{ij}\propto\delta_{ij}$, at both orders,
which in particular excludes a radial anisotropy $\propto n_in_j$; the
quadrupole is then
the coefficient of $-2P_2(\cos\Theta)/R^3$ in $-g_{tt}$. In
Appendix~\ref{app:M2} we show, symbolically in both $p$ and
$\sinh^2\!\alpha$, that these conditions can all be met, that they are
mutually consistent (the system is overdetermined), that the gauge
coefficients they fix are those of the seed, and that the quadrupole readout
depends on only two of them; the same
chart reproduces the exact Geroch--Hansen values $M_2=-m^3q^2$ for vacuum Kerr
and Eq.~\eqref{eq:seedmoments} for vacuum TS$_2$ (cf.\ \cite{Pappas:2012ns}
for the importance of such gauge fixing).

Applying this extraction to the charged metric \eqref{eq:KKTSmetric}, with
both the seed parameter $p$ and the boost parameter $\sinh^2\!\alpha$ kept
symbolic, we find --- for the TS$_2$ seed \emph{and} for the Kerr seed --- the
exact scaling law (derived in Appendix~\ref{app:M2})
\be \label{eq:M2law}
M_2(\alpha)=\frac{1+\cosh^2\!\alpha}{2}\;M_2^{(0)}
=\frac{M}{m}\;M_2^{(0)}\,,
\ee
where $M_2^{(0)}$ is the Geroch--Hansen quadrupole of the vacuum seed; that
is, the invariant mass quadrupole scales under the boost exactly like the
mass. The mechanism is transparent (Appendix~\ref{app:reduction}): in an
ACMC chart the $1/R^2$ term of the seed potential $U=1-f$ has no
$\ell=2$ part, so the nonlinear terms of $-g_{tt}=(1-U)(1+U\sinh^2\!\alpha)^{-1/2}$
cannot feed the $P_2/R^3$ coefficient, the linear coefficient is exactly
$M/m$, and the ACMC chart of the seed remains ACMC after the boost.
Explicitly,
\be \label{eq:M2explicit}
M_2^{\rm TS_2KK}=-\Bigl(q^2+\frac{p^2}{4}\Bigr)m^2 M,
\qquad
M_2^{\rm KK\mbox{-}Kerr}=-q^2\,m^2 M ,
\ee
so the $\delta=2$ quadrupole excess over the KK--Kerr solution with the same
seed parameters,
\be \label{eq:DM2law}
\Delta M_2(\alpha)=-\frac{p^2}{4}\,m^2 M ,
\ee
persists at every boost. Equivalently, in terms of the physical charges, the
reduced quadrupole
\be \label{eq:qhat}
\hat q \equiv -\frac{M_2\,M}{J^2}
=\Bigl(1+\frac{p^2}{4q^2}\Bigr)\frac{(1+\cosh^2\!\alpha)^2}{4\cosh^2\!\alpha}
\ee
exceeds the KK--Kerr value by the boost-\emph{independent} factor
$1+p^2/4q^2$; for Kerr this factor is unity and $\hat q\to1$ as $\alpha\to0$.
Given the causal pathologies of the geometry, we present this ratio as a
particularly clean \emph{asymptotic diagnostic} of the $\delta=2$
deformation rather than as an observable signature.
Since $m=M-\Sigma/\sqrt{3}$ by
\eqref{eq:Mcharge} and \eqref{eq:Scharge}, the seed mass in the law
\eqref{eq:M2law} may be eliminated in favour of the asymptotic ADM and scalar
charges: $M_2\,(M-\Sigma/\sqrt{3})=M\,M_2^{(0)}$.

	\subsection{The singular rod, the two horizons, and the distribution of charge}
	\label{subsec:horizon}
	
The Weyl segment $\mathcal{S}=\{x=1,\ |y|<1\}$ ($\rho=0$, $|z|<\kappa$) of the
seed is \emph{not} a horizon. Kodama and Hikida \cite{Kodama:2003ch} showed
that its induced geometry is Lorentzian, that it is a conical (rod-like)
singularity, and that the degenerate Killing horizons of TS$_2$ are the two
endpoints $\mathcal{H}^\pm=\{x=1,\ y=\pm1\}$, each of which becomes a
two-sphere (with a conical defect at one pole) in suitable regular coordinates.
Gegenberg \emph{et al.}\ \cite{Gegenberg:2010yx} computed the Komar charges of
these objects: the ring singularity carries neither mass nor angular momentum
(as already found in \cite{Kodama:2003ch}), the rod carries all of the mass
and more than all of the angular momentum, $M_{\mathcal{S}}=m$ (Manko
\cite{Manko:2012ts} showed that this mass is not distributed with one sign
along the rod; see below) and
$J_{\mathcal{S}}=J_0-2J_\pm=\frac{2-p}{2(1-p)}J_0>J_0$,\footnote{\label{fn:Jsign}The sign of
$J_\pm$ requires care. Evaluated with the outward orientation used for the
total angular momentum (i.e.\ on closed surfaces that can be deformed into the
sphere at infinity without reversing their orientation), the Komar angular
momenta of the two horizons are \emph{negative}: the horizons counter-rotate
with respect to $J_0$, as found by Manko \cite{Manko:2012ts}, whose
$J_{\rm pole}=-\kappa^2(1+p)/(pq)$ is \eqref{eq:seedhorizon}. This follows
without any orientation ambiguity from Tomimatsu's formulas for the Komar
charges of a rod on which $\omega$ is constant
\cite{Tomimatsu:1983,Tomimatsu:1984}, which hold for singular rods as well as
for horizons \cite{Clement:2019ihb}. With the twist potential
$\chi={\rm Im}\,\mathcal E$, normalized by
$\partial_x\omega=-\kappa(1-y^2)\partial_y\chi/f^2$, they give for the rod
$M_{\mathcal S}=-\tfrac14\omega_{\mathcal S}\bigl[\chi(1,y)\bigr]_{y=-1}^{y=1}$ and
$J_{\mathcal S}=\tfrac12\omega_{\mathcal S}(M_{\mathcal S}-\kappa)$, where
$\omega_{\mathcal S}=2m(1+p)/q$ is the constant value \eqref{eq:rodvalues} of
$\omega$ on the rod. At $x=1$ the potential \eqref{eq:xiTS2} reduces to
$\xi=p-iq(1+y^2)/(2y)$, so that
$\chi(1,y)=-4qy(1+y^2)/[4y^2(1+p)^2+q^2(1+y^2)^2]$ and $\chi(1,\pm1)=\mp q/(1+p)$.
Hence $M_{\mathcal S}=m$ and $J_{\mathcal S}=\frac{2-p}{2(1-p)}J_0$, and since the Komar
two-form of the vacuum seed is closed and the ring carries no Komar charge,
$2J_\pm=J_0-J_{\mathcal S}=-pJ_0/(2(1-p))$. The magnitude
$|J_\pm|=pJ_0/(4(1-p))$ agrees with Eq.~(2.27) of \cite{Gegenberg:2010yx},
which quotes it with a positive sign; their Eq.~(2.28),
$J_{\mathcal{S}}=(1-p/2)J_0$, agrees with neither sign choice.} while each
horizon has
\be \label{eq:seedhorizon}
M_\pm=0,\qquad
J_\pm=-\frac{p}{4(1-p)}\,J_0=-\frac{m^2p}{4}\sqrt{\frac{1+p}{1-p}},\qquad
A_\pm=4\pi r_0^2=2\pi m^2(1+p),
\ee
so that the total horizon area is half that of the Kerr black hole with the
same $(m,J_0)$. In this subsection we determine how the boost acts on each of
these objects. The results are exact and, as for the asymptotic charges, mostly
seed-independent.

\subsubsection*{The rod}

On $\mathcal{S}$ the seed functions evaluate to
\be \label{eq:rodvalues}
f(1,y)=\frac{(1-p)(1-y^2)^2}{(1-p)(1-y^2)^2+8y^2}\in(0,1],
\qquad
e^{2\gamma}(1,y)=\frac{q^4}{p^4},
\qquad
\omega(1,y)=\frac{4\kappa(1+p)}{pq},
\ee
the last being $y$-independent. The combination
$\partial_t+\Omega_{\mathcal{S}}\partial_\phi$ with
\be \label{eq:OmegaS}
\Omega_{\mathcal{S}}=\frac{1}{\omega(1,y)}=\frac{q}{2m(1+p)}
=\frac{p}{4\kappa}\sqrt{\frac{1-p}{1+p}}
\ee
has vanishing norm on the rod; this is the quantity denoted
$\Omega_{\mathcal{S}}$ in \cite{Gegenberg:2010yx} (their Eq.~(2.42)), and it
characterizes the helical structure of the singular rod, not a horizon
angular velocity. Since $f(1,y)>0$ away from the endpoints,
$\tilde g_{\phi\phi}(1,y)=-f\omega^2<0$: the orbits of $\partial_\phi$ are
timelike on and adjacent to the rod, which is the causality violation of
\cite{Kodama:2003ch}. For the charged solution the corresponding statements
are obtained from \eqref{eq:WPcharged}: the vanishing Killing combination on
the rod becomes $\partial_t+(\Omega_{\mathcal{S}}/\cosh\alpha)\partial_\phi$,
the rod remains a Lorentzian line with $g'_{\phi\phi}(1,y)=-(f/H)\cosh^2\!
\alpha\,\omega^2<0$, and the CTCs adjacent to it persist for all $\alpha$
(Sec.~\ref{subsec:global}).

\subsubsection*{The horizons}

At the endpoints $\mathcal{H}^\pm$ the seed potential vanishes, $f\to0$, and
hence the boost function takes the \emph{exact} value
\be \label{eq:Hhorizon}
H\big|_{\mathcal{H}^\pm}=\cosh\alpha .
\ee
More precisely, in the bipolar coordinates $(\tilde r,\theta)$ of
\cite{Gegenberg:2010yx}, in which $\mathcal{H}^+$ is the surface $\tilde r=0$,
one has $f=\mathcal{O}(\tilde r^2)$ and $f\omega=\mathcal{O}(\tilde r)$, so
that $H=\cosh\alpha+\mathcal{O}(\tilde r^2)$ and $\partial H=\mathcal{O}(\tilde
r)$. Inserting $f'=f/H$, $\omega'=\cosh\alpha\,\omega$, $\gamma'=\gamma$ into
the Weyl--Papapetrou form \eqref{eq:WP} and using \eqref{eq:Hhorizon},
\be \label{eq:NHscaling}
\begin{gathered}
g'_{\mu\nu}=\cosh\alpha\;\hat g_{\mu\nu}\,\bigl[1+\mathcal{O}(\tilde r^2)\bigr]
\quad\text{(no sum)},\\
d\hat s^2\equiv-f\,(d\hat t-\omega\,d\phi)^2
+f^{-1}\bigl(e^{2\gamma}(d\rho^2+dz^2)+\rho^2d\phi^2\bigr),
\qquad \hat t\equiv\frac{t}{\cosh\alpha},
\end{gathered}
\ee
component by component in the coordinates $(\hat t,\tilde r,\theta,\phi)$:
each component of the charged metric equals $\cosh\alpha$ times the
corresponding component of the seed (with rescaled time) up to a
\emph{relative} correction of order $\tilde r^2$. (The correction is relative,
not absolute: components such as $g_{\tilde r\tilde r}$ diverge at $\tilde r=0$.)
Consequently the near-horizon scaling limit of \cite{Gegenberg:2010yx} applied
to the charged metric gives exactly $\cosh\alpha$ times the near-horizon limit
of the seed: the charged horizon geometry is a \emph{constant conformal
rescaling} of the seed's, with a rescaled time. Everything about the horizons
follows from \eqref{eq:NHscaling}. We stress what these objects are: degenerate
Killing horizons whose spatial sections are two-spheres with a conical defect
at the pole adjoining the rod, inherited unchanged from the seed (a constant
rescaling preserves the deficit angle); they are not smooth, isolated
black-hole horizons, and their properties below should be read in that light.

\emph{Degeneracy and non-rotation are preserved.} A constant rescaling of the
metric and of the Killing time does not change the location of a Killing
horizon or the vanishing of its surface gravity, and the null generator is
still proportional to $\partial_t$: $T_\pm=0$ and $\Omega_\pm=0$ for every
$\alpha$, even though the solution carries $J=\cosh\alpha\,J_0\neq0$.
$\Omega_\pm=0$ is a statement about the horizon generator only; the
near-horizon geometry itself is rotating (it carries the Komar angular
momentum $J_\pm\neq0$ below and has $g'_{t\phi}\neq0$ off the horizon). The
near-horizon geometry is $\cosh\alpha$ times the near-horizon TS$_2$ geometry
of \cite{Gegenberg:2010yx}, which belongs to the general class of vacuum
near-horizon extremal spinning metrics; the matter fields are constant there
(see below), so the near-horizon limit of the charged solution is an exact
EMD solution with constant scalars --- the natural starting point for a KK
Kerr/CFT analysis along the lines of \cite{Siahaan:2025mkk,Siahaan:2024ilq}.

\emph{Areas.} The induced two-metric on $\mathcal{H}^\pm$ is exactly
$\cosh\alpha$ times the seed's: in $g'_{\phi\phi}$ the term
$(f/H)\cosh^2\!\alpha\,\omega^2$ tends to $\cosh\alpha\,f\omega^2$, and
$f\omega^2$ is finite on the horizon. Hence
\be \label{eq:areacharged}
A_\pm(\alpha)=\cosh\alpha\;A_\pm=\cosh\alpha\;2\pi m^2(1+p),
\ee
the same scaling as for the boosted Kerr black hole, and the conical defect at
the pole is unchanged. The dilaton is constant on each horizon,
$\Phi|_{\mathcal{H}^\pm}=\tfrac{\sqrt3}{2}\ln\cosh\alpha$.

\emph{Electric potential.} Since $f\omega\to0$ on the horizon,
$A_\phi=\sinh\alpha\,f\omega/2H^2\to0$ --- the gauge \eqref{eq:At}--\eqref{eq:Aphi}
is regular there --- and with $\Omega_\pm=0$ the corotating potential is
\be \label{eq:PhiH}
\Phi_e\equiv\bigl(A_t+\Omega_\pm A_\phi\bigr)\big|_{\mathcal{H}^\pm}
=\frac{\cosh\alpha\sinh\alpha}{2\cosh^2\!\alpha}=\frac12\tanh\alpha ,
\ee
constant on both horizons: the zeroth law for the electric potential holds,
with the same value as for all boosted KK black holes.

\emph{Komar charges.} Under $g\to c\,g$ with constant $c$ the Komar two-form
of a fixed Killing vector scales by $c$; combined with the time rescaling in
\eqref{eq:NHscaling} and the fact that first derivatives of $H$ vanish on
$\tilde r=0$, this gives, for the Komar angular momentum and mass of each
horizon,
\be \label{eq:komarcharged}
J_\pm(\alpha)=\cosh\alpha\;J_\pm=-\cosh\alpha\;\frac{m^2p}{4}\sqrt{\frac{1+p}{1-p}},
\qquad
M_\pm(\alpha)=M_\pm=0 .
\ee
Here $M_\pm$ denotes the Komar mass of $\partial_t$, not every possible notion
of horizon mass. Its vanishing is in fact forced: for any Killing horizon the
Komar mass of $\partial_t$ equals $\kappa A/4\pi+2\Omega_H J_H$, and here
$\kappa=\Omega_H=0$.

\emph{Electric charge: an exact flux identity.} Where the electric charge
resides is settled by an identity that holds throughout the spacetime, not only
on the horizons. The KK gauge field is the Komar potential of the KK Killing
vector $\partial_{z'}=\sinh\alpha\,\partial_t+\cosh\alpha\,\partial_z$ of the
(Ricci-flat) boosted five-metric. Since $\partial_z$ is covariantly constant
its Komar two-form vanishes identically, and what remains is $\sinh\alpha$
times the Komar two-form of $\partial_t$ of the seed, integrated over the
tilted KK circle, which supplies a further factor $\cosh\alpha$. The outcome is
an identity between the pull-backs of two closed two-forms to every
hypersurface $t={\rm const}$,
\be \label{eq:fluxidentity}
e^{2\sqrt3\Phi}\,{\star F}\big|_{t={\rm const}}
=\frac{\sinh\alpha\cosh\alpha}{2}\;\tilde\star\,d\tilde\xi_{(t)}\big|_{t={\rm const}},
\ee
which we have verified, component by component, by direct symbolic computation
for an arbitrary Weyl--Papapetrou seed --- arbitrary functions $f$, $\omega$,
$\gamma$ --- and arbitrary $\alpha$ (tilde: seed quantities, $\tilde\xi_{(t)}$
the one-form dual to $\partial_t$ of the seed); it uses nothing but the boost
\eqref{eq:KKTSmetric}. The $d\rho\wedge dz$, $d\rho\wedge d\phi$,
$dz\wedge d\phi$ and also the $dt\wedge d\phi$ components agree, while the
$dt\wedge d\rho$ and $dt\wedge dz$ components differ by
$\mp(\cosh\alpha-1)\sinh\alpha\,f^2\partial_{z,\rho}\omega/(2\rho)$, so
\eqref{eq:fluxidentity} is not an identity of four-dimensional two-forms. It
is all that is needed: both forms are closed in the source-free region and
invariant under $\partial_t$, so any closed two-surface can be moved along
$\partial_t$ into a slice $t={\rm const}$ without changing either flux.
With our sign convention $A_t\to+Q/r$ (footnote~\ref{fn:signs}) the electric
charge enclosed by a closed two-surface $S$, with the outward orientation, is
$Q(S)=-\frac{1}{4\pi}\oint_S e^{2\sqrt3\Phi}\star F$, and the Komar mass of the
seed enclosed by $S$ is $M^{(0)}_{\rm K}(S)=-\frac{1}{8\pi}\oint_S\tilde\star\,
d\tilde\xi_{(t)}$; \eqref{eq:fluxidentity} then reads
\be \label{eq:fluxQM}
Q(S)=\sinh\alpha\cosh\alpha\;M^{(0)}_{\rm K}(S)
\ee
for \emph{every} closed two-surface $S$ in the physical domain. Two checks fix
the relative sign independently: at infinity \eqref{eq:fluxQM} reproduces
$Q=m\sinh\alpha\cosh\alpha$, and for small boosts \eqref{eq:gauge} gives
$A=\tfrac{\alpha}{2}(dt+\tilde\xi_{(t)})+\mathcal{O}(\alpha^3)$, hence
$F=\tfrac{\alpha}{2}d\tilde\xi_{(t)}$ and $Q(S)=\alpha M^{(0)}_{\rm K}(S)+
\mathcal{O}(\alpha^3)$. The content of \eqref{eq:fluxQM} is that the electric
charge of the solution is distributed exactly as the Komar mass of the vacuum
seed. The right-hand side is a vacuum Komar integral: the seed's Komar two-form
is closed wherever the seed is Ricci-flat, i.e.\ everywhere except on the rod,
the horizons and the ring singularity, so it depends only on which of these
seed sources $S$ encloses, and the charge bookkeeping is fixed by the seed
Komar masses quoted at the beginning of this subsection:
\begin{itemize}
\item[(i)] the horizons carry no charge,
\be \label{eq:Qpm}
Q_\pm=-\frac{1}{4\pi}\oint_{\mathcal{H}^\pm}e^{2\sqrt3\Phi}\,{\star F}
=\sinh\alpha\cosh\alpha\,M_\pm=0 ,
\ee
in agreement with the direct near-horizon estimate (with
$A_t=\tfrac12\tanh\alpha+\mathcal{O}(\tilde r^2)$, $A_\phi=\mathcal{O}(\tilde
r)$, the flux through $\tilde r=\epsilon$ vanishes as $\epsilon\to0$);
\item[(ii)] the excised lens $H^2\le0$ of Sec.~\ref{subsec:global}, whose
boundary $H^2=0$ is a new inner boundary of the four-dimensional domain, and
the ring singularity, which lies on that boundary (every level set of $f$
reaches the ring, Sec.~\ref{subsec:global}), carry no charge either. Since the
two cannot be separated, their charge is defined by enclosing surfaces. The
closure of the excised set $\mathcal{E}_\alpha=\{f\ge\coth^2\!\alpha\}$ is
compact and bounded away from the axis ($f<1$ there, since
$f(x,\pm1)=\mu^2/(\mu^2+\mu\nu)$ with $\mu,\nu>0$), from the rod ($f\le1$),
from the horizons ($f=0$) and from infinity ($f\to1$), so there are closed
surfaces $S$ in the physical domain that enclose $\mathcal{E}_\alpha$ (and with
it the ring) but neither the rod nor the horizons. In the seed such an $S$
encloses no source but the ring; since the seed Komar form is closed in the
region it bounds minus the ring, $M^{(0)}_{\rm K}(S)$ equals the Komar mass of
an arbitrarily small surface around the ring, which vanishes
\cite{Kodama:2003ch,Gegenberg:2010yx}; hence
\be \label{eq:Qlens}
Q_{\rm lens+ring}\equiv Q(S)
=\sinh\alpha\cosh\alpha\,M^{(0)}_{\rm K}({\rm ring})=0 ;
\ee
\item[(iii)] consequently, by Gauss' law in the source-free physical domain,
the entire charge is sourced by the rod: $Q_{\mathcal{S}}=Q$. The identity
\eqref{eq:fluxQM} says more: the charge is distributed along the rod exactly
as the seed Komar mass, and that distribution is not of one sign. Manko
\cite{Manko:2012ts} showed that the Komar mass $M(z)$ of the seed enclosed by
a thin surface around the portion $|z'|<z$ of the rod increases up to
$z_{\max}$ and decreases beyond it, so that the inner segment carries
$M_{\mathcal{S}}^{+}=m/q$ and the two outer pieces carry
$M_{\mathcal{S}}^{-}=-m(1-q)/q$. In terms of the twist potential on the rod
(footnote~\ref{fn:Jsign}) his mass function is
$M(z)=-\tfrac12\omega_{\mathcal S}\,\chi(1,z/\kappa)$; writing
$\chi(1,y)=-\frac{2}{1+p}\,\frac{u}{1+u^2}$ with $u=2y(1+p)/(q(1+y^2))$, its
extremum $\chi=-1/(1+p)$ at $u=1$ gives $M_{\mathcal S}^+=m/q$ and lies at
\be \label{eq:zmax}
\frac{z_{\max}}{\kappa}=\frac{\sqrt{p}-\sqrt{1-q}}{\sqrt{p}+\sqrt{1-q}}
\qquad(=2-\sqrt3\ \text{at}\ p=\tfrac35).
\ee
By \eqref{eq:fluxQM} the charged rod therefore carries
$Q_{\mathcal{S}}^{+}=Q/q>Q$ on $|z|<z_{\max}$ and
$Q_{\mathcal{S}}^{-}=-Q(1-q)/q$ on $z_{\max}<|z|<\kappa$: the KK charge of
the rod is of both signs, with the negative charge sitting next to the
horizons.
\end{itemize}

\subsubsection*{A boost-invariant global identity}

Combining \eqref{eq:Mcharge}, \eqref{eq:Qcharge} and \eqref{eq:PhiH},
\be \label{eq:smarrdefect}
M-\Phi_e\,Q
=\frac{m}{2}\bigl(1+\cosh^2\!\alpha\bigr)-\frac{m}{2}\sinh^2\!\alpha
= m ,
\ee
independently of $\alpha$. We present \eqref{eq:smarrdefect} as a
\emph{boost-invariant global identity} among the asymptotic charges $M$, $Q$
and the horizon potential $\Phi_e$, and not as a Smarr relation: for a regular
KK black hole one has $M=2TS+2\Omega_HJ+\Phi_eQ$, whereas here the horizons
are extremal ($T_\pm=0$), non-rotating ($\Omega_\pm=0$) and uncharged
($Q_\pm=0$) and contribute nothing to such a balance. The invariant $m$ on the
right-hand side is the seed mass, which by \eqref{eq:fluxQM} is also
$Q/(\sinh\alpha\cosh\alpha)$. We deliberately refrain from reading
\eqref{eq:smarrdefect} as a statement about where the \emph{mass} of the
charged solution resides: in a matter-coupled spacetime the Komar two-form of
$\partial_t$ is not closed in the bulk, so the Komar mass of the charged
solution is surface-dependent and no localization on the rod is implied. The
only localization statements we make are those that follow from closed
two-forms --- the Maxwell flux, through \eqref{eq:fluxidentity}, and the Komar
charges of the vacuum seed. We note that \eqref{eq:smarrdefect} is
seed-independent and that for the Kerr seed, where the horizon is regular, it
combines with the Smarr relation of the KK--Kerr black hole into
$2TS+2\Omega_HJ=m$, the boost-invariant form of the familiar Kerr balance
$2TS=mp$.

	\subsection{Ergoregion, new singularities, and causal structure}
	\label{subsec:global}

\subsubsection*{Ergosurface}

Since $g_{tt}=-f/H$ and $H^2\to\cosh^2\!\alpha$ as $f\to0$, the ergosurface of
the charged solution sits at exactly the seed locus $f=0$ for \emph{every}
boost; only its intrinsic geometry is deformed by the factor $H$. For TS$_2$
the ergoregion does not hug a horizon as in Kerr: $f(1,y)>0$ on the rod by
\eqref{eq:rodvalues}, so the ergosurface meets the rod only at its two
endpoints --- that is, precisely at the horizons $\mathcal{H}^\pm$, where
$f=0$ --- and the ergoregion proper is the shaded region of
Fig.~\ref{fig:structure}(a), detached from the rod elsewhere.

\subsubsection*{A new curvature singularity at $H^2=0$}

The genuinely new global feature of the charged solution is the locus
\be \label{eq:H0locus}
H^2=0
\quad\Longleftrightarrow\quad
f=\coth^2\!\alpha>1 .
\ee
The seed possesses a region with $f>1$, and its existence can be established
exactly. On the equator $y=0$ the polynomials of \eqref{eq:ABCform} factorize,
$A(x,0)=N(x)D(x)$ and $B(x,0)=D(x)^2$ with
$N=p^2x^4-2px^3+2px-1$ and $D=p^2x^4+2px^3-2px-1$, so that
\be \label{eq:feq}
f(x,0)=\frac{N(x)}{D(x)},
\qquad
f(x,0)-1=\frac{-4px\,(x^2-1)}{p^2x^4+2px^3-2px-1}.
\ee
The ring singularity $B=0$ is the root $x_s>1$ of $D$, equivalently of
\be \label{eq:ringlocus}
p\,(x^2-1)\,(px^2+2x+p)=q^2 ,
\qquad y=0
\ee
(e.g.\ $x_s\simeq1.1368$ for $p=0.6$); since $D(1)=p^2-1<0$ and
$D'(x)=2p(2px^3+3x^2-1)>0$ for $x\ge1$, it is the only root of $D$ with
$x>1$. Hence on $1<x<x_s$ the denominator of \eqref{eq:feq} is
negative and approaches zero from below: $f(x,0)-1>0$, with $f(1,0)=1$ exactly
and $f\to+\infty$ as $x\to x_s^-$. By continuity, $f=\coth^2\!\alpha>1$ has at
least one equatorial root for every $\alpha\neq0$ throughout the
\emph{rotating} family $0<p<1$; no monotonicity assumption is needed. This
proves an \emph{obstruction}: since $f$ is unbounded above near the ring, no
nonzero boost leaves the KK direction spacelike throughout the seed domain. More
generally, by \eqref{eq:HTS} a boost of a vacuum seed defines a spacelike
reduction everywhere if and only if $f<\coth^2\!\alpha$ throughout the seed
domain. For the Kerr seed $f=1-2mr/(r^2+j^2m^2\cos^2\!\theta)\le1$ for $r\ge0$
($j=J_0/m^2$), so
every boost is admissible; for TS$_2$, $f$ is unbounded above and none is.

What is proved, then, is that the excised set $\{f\ge\coth^2\!\alpha\}$ is
nonempty for every $\alpha\neq0$ and $0<p<1$, and --- because $f<1$ on the axis,
$f\le1$ on the rod, $f=0$ on the horizons and $f\to1$ at infinity --- that its
closure is compact and touches neither the axis, the rod, the horizons nor
infinity. It contains the ring, which lies on its boundary: at the ring $A$ and
$B$ vanish together and $f$ takes all real values in every neighbourhood of it
(on a small circle of radius $\delta$ around the ring in the $(x,y)$ plane, $f$
ranges between values of order $\mp1/\delta$), so every level set of $f$, and
in particular the boundary $H^2=0$, reaches the ring. The sets are nested,
$\{f\ge\coth^2\!\alpha\}\subset\{f\ge\coth^2\!\alpha'\}$ for $\alpha<\alpha'$;
Fig.~\ref{fig:structure}(a) shows them for three values of $\alpha$, and we
refer to the excised set as the lens, without claiming that it is connected in
general. Rotation is essential: in the static limit $p=1$
(Sec.~\ref{subsec:static}) one has $f=((x-1)/(x+1))^2<1$ everywhere and no
such region exists. In the extreme-rotation limit $p\to0$ one must say what is
held fixed: the prolate coordinate of the ring diverges, $x_s\simeq(2p)^{-1/3}$,
but at fixed seed mass its Weyl radius $\rho_s=\kappa\sqrt{x_s^2-1}\simeq
2^{-4/3}p^{2/3}m$ tends to zero, together with the rod ($|z|<\kappa=mp/2$), so
that the lens, pinched between the two, shrinks onto the origin. As
$\alpha\to0^+$ the lens shrinks onto the ring singularity; as $\alpha\to\infty$
it approaches the level set $f=1$, which meets the rod at its equatorial point
$y=0$. It never reaches the horizons $\mathcal{H}^\pm$, where $f=0$ and
$H=\cosh\alpha$ exactly (Sec.~\ref{subsec:horizon}). In the region $f>\coth^2\!\alpha$ one has
$H^2<0$ and the fields \eqref{eq:KKTSmetric}--\eqref{eq:dilatonTS} cease to
be real: this lens is excised from the charged spacetime. The physical domain
of the solution is therefore
\be \label{eq:domain}
x\geq1,\qquad -1\leq y\leq1,\qquad H^2=\cosh^2\!\alpha-f\sinh^2\!\alpha>0,
\ee
and the fields and their curvature invariants are defined only there.

The surface \eqref{eq:H0locus} is a genuine curvature singularity. Taking the
trace of \eqref{eq:Rmn} (the Maxwell stress is traceless) gives
$R=2\,\partial_\mu\Phi\,\partial^\mu\Phi$, and using
$\partial H/\partial x=-\sinh^2\!\alpha\,f_{,x}/(2H)$ one finds the exact
expression
\be \label{eq:Rblowup}
R=\frac{3\sinh^4\!\alpha\;f}
{8\kappa^2 e^{2\gamma}(x^2-y^2)}\;
\frac{(x^2-1)f_{,x}^{\,2}+(1-y^2)f_{,y}^{\,2}}{H^{5}} ,
\ee
which diverges as $H^{-5}$ wherever $\nabla f\neq0$ on the level set --- a
naked curvature singularity of the four-dimensional Einstein-frame geometry,
absent in the vacuum seed and present for arbitrarily small charge; $RH^5$
tends to the finite, positive value of the prefactor in \eqref{eq:Rblowup}
(Fig.~\ref{fig:structure}(c) shows $R$ along the equator). By
contrast, at the ergosurface $H$ remains finite ($H=\cosh\alpha$), and the
dilaton and Maxwell fields are regular there.

\begin{figure}[tp]
\centering
\includegraphics[width=0.84\textwidth]{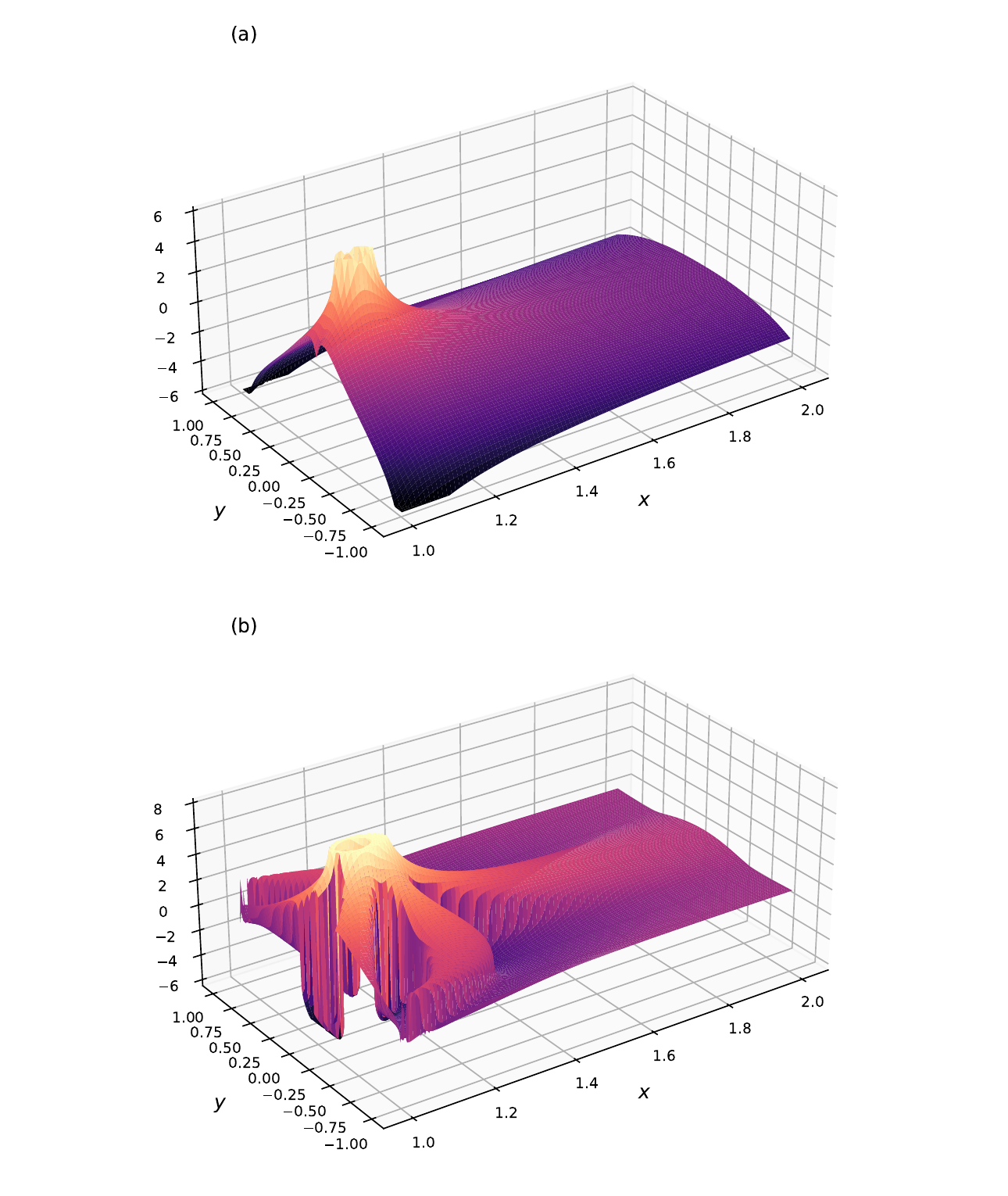}
\caption{Curvature invariants of the charged solution for $p=0.6$,
$\kappa=0.3$, $\alpha=1$, plotted over $1\le x\le2$, $|y|\le1$ on the physical
domain \eqref{eq:domain}. (a) $\log_{10}R$, where $R=2(\partial\Phi)^2$ is
evaluated from the exact expression \eqref{eq:Rblowup}; clipped to $[-6,6]$.
(b) The Kretschmann scalar on the signed logarithmic scale
$\mathrm{sgn}(K)\log_{10}(1+|K|)$, computed from the Riemann tensor of
\eqref{eq:KKTSmetric}; clipped to $[-8,8]$. Both diverge on the lens boundary
$H^2=0$, which bounds the hole in each surface; $K$ diverges also at the ring
singularity of the seed.}
\label{fig:curv}
\end{figure}

The five-dimensional origin of this singularity clarifies its nature. The
boost is a coordinate transformation of the covering space
${\rm TS}_2\times\mathbb{R}$, so the five-dimensional geometry acquires no new
local curvature singularity away from the seed's own singular set (the ring
and the rod, which it inherits); the KK solution is obtained by imposing the
periodicity along the \emph{new} direction $z'$, and is therefore globally a
different spacetime from ${\rm TS}_2\times S^1$. What the boost changes is the
Killing direction along which one reduces. The norm of that direction is
$G_{z'z'}=H^2$ (cf.\ \eqref{eq:boosted5d}), so the locus \eqref{eq:H0locus} is
where the KK circle becomes \emph{null}, and beyond it the compact direction
is timelike. The four-dimensional Einstein-frame metric $g_{\mu\nu}=H\,
G_{\mu\nu}^{(4)}$ and the dilaton $\Phi=\tfrac{\sqrt3}{2}\ln H$ are singular
there because the reduction itself degenerates. The new singularity is thus a
\emph{KK-reduction singularity}: a four-dimensional curvature singularity
generated by the degeneration of the reduction circle inside the $f>1$ region
that only the rotating TS$_2$ seed provides, with no five-dimensional
counterpart. Equation \eqref{eq:Rblowup} also controls the far field: with
$f=1-2m/r+\mathcal{O}(r^{-2})$ and $H=1+m\sinh^2\!\alpha/r+\mathcal{O}(r^{-2})$
it gives
\be \label{eq:Rasym}
R=\frac{3m^2\sinh^4\!\alpha}{2\,r^4}+\mathcal{O}(r^{-5})
=\frac{2\Sigma^2}{r^4}+\mathcal{O}(r^{-5}),
\ee
which vanishes in the vacuum limit and provides an independent check of the
dilaton charge \eqref{eq:Scharge}. (For large boosts the asymptotic regime
sets in only for $r\gg m\sinh^2\!\alpha$, since $H^2-1\simeq2m\sinh^2\!\alpha/r$
must be small.) The Kretschmann invariant, as for any asymptotically flat
solution with mass $M$, begins as $48M^2/r^6$, the Maxwell and dilaton
contributions entering at $\mathcal{O}(r^{-7})$. Both invariants are shown in
Fig.~\ref{fig:curv} on the physical domain. The Ricci scalar, computed from
the closed form \eqref{eq:Rblowup}, is a smooth positive surface whose only
feature is the ridge along the lens boundary, where it grows as $H^{-5}$; the
Kretschmann scalar, computed independently from the full Riemann tensor of
\eqref{eq:KKTSmetric}, diverges both at the seed's ring singularity and along
the same lens boundary, and changes sign across the domain as it does for
Kerr-type vacua.

\subsubsection*{Closed timelike curves}

\begin{figure}[tp]
\centering
\includegraphics[width=0.9\textwidth]{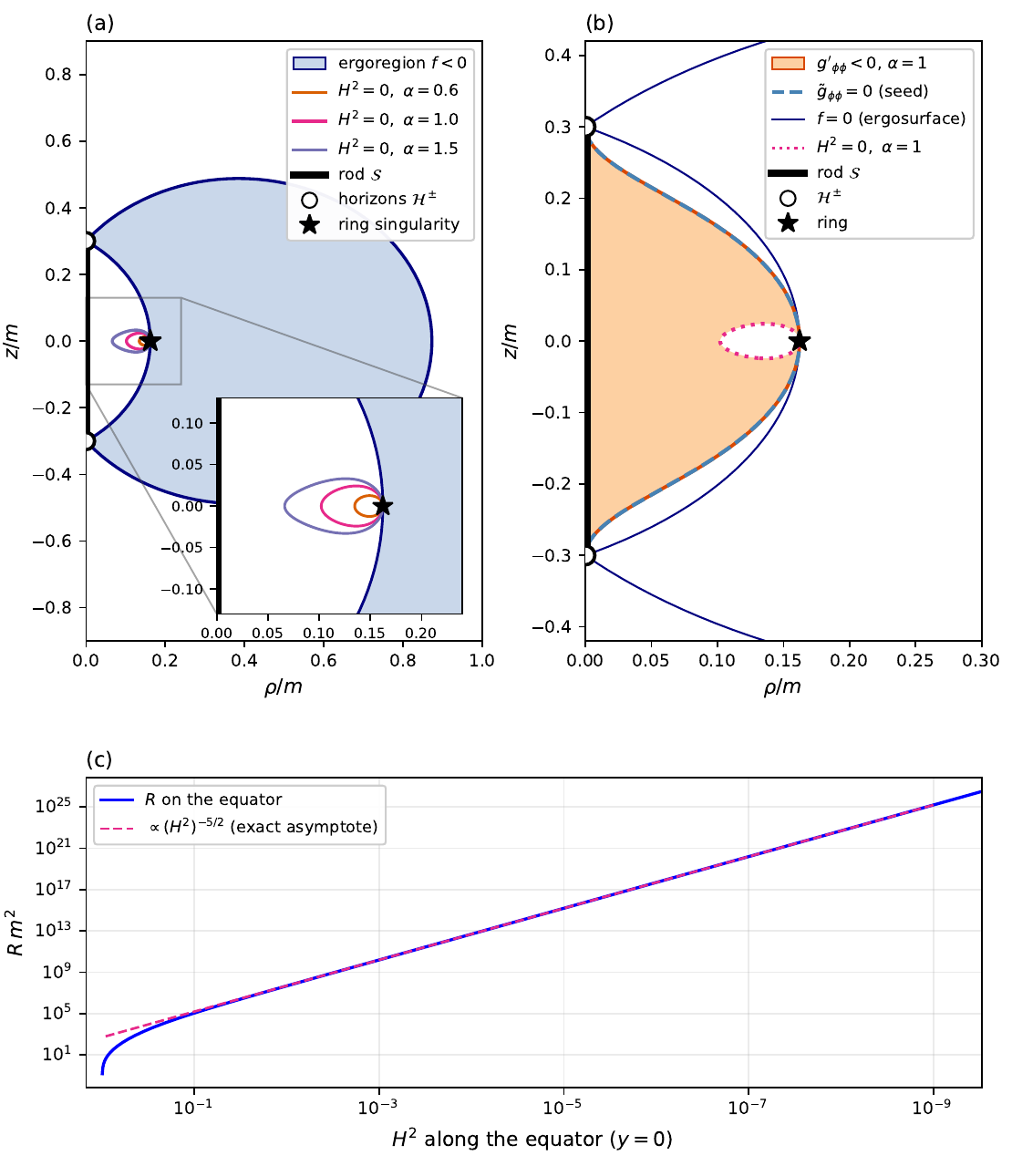}
\caption{Global structure of the charged TS$_2$ solution for $p=0.6$ in the
Weyl half-plane (lengths in units of $m$; the $\rho$ and $z$ scales differ).
(a) The ergoregion $f<0$ (shaded) with its boundary $f=0$, which is
boost-invariant; the $H^2=0$ reduction singularities $f=\coth^2\!\alpha$ for
$\alpha=0.6,\,1.0,\,1.5$; the singular rod $\mathcal{S}$ at $\rho=0$,
$|z|<\kappa$ (thick line), the two degenerate horizons $\mathcal{H}^\pm$ at its
endpoints (open circles), and the ring singularity of the seed (star). The
inset enlarges the region $0\le\rho\le0.24\,m$, $|z|\le0.13\,m$ around the ring,
where the lenses attached to the ring are visible. The lenses are nested, all
have the ring on their boundary, grow with $\alpha$ towards the level set
$f=1$, which touches the rod at its midpoint, and stay away from the horizons.
(b) Enlargement of the rod neighbourhood ($0\le\rho\le0.3\,m$): the
causality-violating region $g_{\phi\phi}<0$ of the charged solution at
$\alpha=1$ (shaded, solid boundary) and the boundary $\tilde g_{\phi\phi}=0$ of
the seed's (dashed, almost coincident). The region fills the layer between the
rod and a curve just inside the ergosurface $f=0$ (thin line) and is bounded by
the horizons; the unshaded region attached to the ring is the excised lens
$H^2\le0$ (dotted boundary). (c) The Ricci scalar \eqref{eq:Rblowup}, in units
of $m^{-2}$, along the equator $y=0$ against $H^2$ (log--log, $H^2\to0$ to the
right) at $\alpha=1$ (solid), with its exact asymptote
$\propto(H^2)^{-5/2}=H^{-5}$ from \eqref{eq:Rblowup} (dashed).}
\label{fig:structure}
\end{figure}

The spatial components of the reduced metric follow directly from the
five-dimensional form \eqref{eq:boosted5d}:
\be \label{eq:gijrel}
g_{ij}=H\,\tilde g_{ij}-\frac{\sinh^2\!\alpha}{H}\,\tilde g_{ti}\,\tilde g_{tj},
\ee
so that $g_{xx}=H\tilde g_{xx}$, $g_{yy}=H\tilde g_{yy}$ exactly, while
\be \label{eq:ctccond}
g'_{\phi\phi}=H\,\tilde g_{\phi\phi}-\frac{\sinh^2\!\alpha}{H}\,\tilde g_{t\phi}^{\,2}
=\frac{H}{f}\Bigl[\kappa^2(x^2-1)(1-y^2)
-\frac{\cosh^2\!\alpha}{H^2}\,f^2\omega^2\Bigr].
\ee
On the physical domain $H>0$, so the first form shows at once that
$g'_{\phi\phi}\le H\tilde g_{\phi\phi}$, with equality only where
$\tilde g_{t\phi}=f\omega=0$: every point at which the orbits of
$\partial_\phi$ are timelike in the seed ($\tilde g_{\phi\phi}<0$) and which
survives the excision has them timelike in the charged solution as well, and
the second term can create new CTCs where $\tilde g_{\phi\phi}$ is small and
positive. Charging therefore \emph{strengthens} causality violation pointwise
throughout the physical domain. In particular the CTCs on and adjacent to the
rod persist for all $\alpha$, with $g'_{\phi\phi}(1,y)=-(f/H)\cosh^2\!\alpha
\,\omega^2<0$. For $p=0.6$ the seed CTC region lies in the $f>0$ layer between
the rod and the ring singularity (Fig.~\ref{fig:structure}(b)) and the charged
region contains its surviving part; the only loss is the part of the seed's CTC
region that falls inside the excised lens \eqref{eq:H0locus}, which grows with
$\alpha$. Charging therefore does not cure the causal pathology of the seed; it
partially replaces it by a reduction singularity.

	\section{Conclusions}
	\label{sec:conclusions}

We have constructed the Kaluza--Klein dressing of the \TS{} $\delta=2$
spacetime by the uplift--boost--reduce procedure and analysed its physical
properties. Three results stand out.

First, at the level of the leading asymptotic charges the solution is
indistinguishable from the boosted Kerr black hole: mass, angular momentum,
electric and dilaton charges, magnetic dipole, the KK constraint
\eqref{eq:KKconstraint}, and the gyromagnetic ratio \eqref{eq:gfactor} are
all $\delta$-independent. The seed reveals itself one multipole order deeper.
We computed the Geroch--Hansen moments of the vacuum seed,
Eq.~\eqref{eq:seedmoments}, constructed an ACMC chart for the charged
solution --- which, by Mayerson's generalization of the Geroch--Hansen
formalism, yields its invariant multipoles --- and derived, symbolically in
$p$ and $\sinh^2\!\alpha$, the exact boost-scaling law \eqref{eq:M2law},
$M_2(\alpha)=(M/m)\,M_2^{(0)}$, for both the Kerr and the TS$_2$ seeds. The
$\delta=2$ quadrupole excess \eqref{eq:DM2law} therefore survives at every
boost, and reduced quadrupoles form boost-invariant ratios: a clean
asymptotic diagnostic of the deformation.

Second, we determined what the boost does to the global structure identified
by Kodama and Hikida and by Gegenberg \emph{et al.}: a singular rod carrying
the mass, and two degenerate horizons at its endpoints. On the horizons the
boost function is exactly $H=\cosh\alpha$, and the near-horizon geometry of
the charged solution is a constant rescaling of the seed's. Consequently the
horizons remain degenerate and non-rotating, their areas and Komar angular
momenta scale by $\cosh\alpha$, the electric potential on them is the
constant $\tfrac12\tanh\alpha$ with the dilaton likewise constant. These
statements are seed-independent consequences of $f\to0$ on a degenerate
horizon. Where the charge resides is fixed
by an exact identity: the electric flux through any closed surface is
$\sinh\alpha\cosh\alpha$ times the Komar mass of the vacuum seed enclosed by
it, a four-dimensional transcription of the five-dimensional Komar integral of
the KK Killing vector. Since the horizons and the ring singularity carry no
seed Komar mass, they carry no charge --- including the new inner boundary
$H^2=0$ of the physical domain, on which the ring lies --- and the entire
charge is sourced by the rod. The combination $M-\Phi_eQ=m$ is a
boost-invariant global identity among the asymptotic charges; we do not read
it as a localization of the mass, since Komar integrals are surface-dependent
in the matter-filled bulk.

Third, the ergosurface is pinned at the seed locus $f=0$, while a new
curvature-singular boundary develops on the surface $f=\coth^2\!\alpha$, with
$R\sim H^{-5}$, Eq.~\eqref{eq:Rblowup}. Its five-dimensional origin is
transparent: the boost is a coordinate transformation of the covering space
${\rm TS}_2\times\mathbb{R}$, the periodicity is imposed along the new
direction, and the new locus is where that Killing circle becomes null, so this is a KK-reduction singularity of the
four-dimensional Einstein frame with no five-dimensional counterpart. It
exists for every nonzero boost throughout the rotating family $0<p<1$, because
$f$ is unbounded above near the ring: a boost is an admissible spacelike
reduction everywhere only if $f<\coth^2\!\alpha$ throughout the seed, which
holds for every boost of Kerr ($f\le1$) and for none of TS$_2$. The excised
region is compact, has the ring on its boundary, never reaches the horizons,
and interpolates between the ring ($\alpha\to0$) and the region $f\ge1$
($\alpha\to\infty$). Causality violation is strengthened pointwise by the
charge, $g'_{\phi\phi}\le H\tilde g_{\phi\phi}$, the only CTC region lost being
the part swallowed by the excised lens. In contrast to the Einstein--Maxwell magnetization of TS$_2$, which
can remove the ring singularity \cite{Manko:1999ts}, the KK dressing leaves
the ring in place and adds a singular boundary.

Several directions follow naturally. The boost is only the simplest charging
transformation in the $SL(3,\mathbb{R})$ solution-generating group of the
theory \cite{Dobiasch:1981vh,Breitenlohner:1987dg,Rasheed:1995zv}; a dyonic
TS$_2$ solution, or a Melvin--TS$_2$ spacetime obtained by the KK Harrison
transformation \cite{Yazadjiev:2013qna,Siahaan:2026mzv}, would deform the
multipole structure more radically and would show whether the dilaton
preserves the regularizing effect of a strong magnetic field found in
\cite{Manko:1999ts}. The near-horizon geometry derived here --- a constant
rescaling of the near-horizon TS$_2$ metric with constant Maxwell potential
and dilaton --- is the natural starting point for a Kerr/CFT description of
the charged horizons, extending the vacuum analysis of
\cite{Gegenberg:2010yx} and the magnetized and accelerating KK analyses of
\cite{Siahaan:2025mkk,Siahaan:2024ilq}; the horizon charges $J_\pm(\alpha)$
and $Q_\pm=0$ obtained above are the inputs such an analysis needs. The
argument of Appendix~\ref{app:reduction} establishes \eqref{eq:M2law}
for every Weyl--Papapetrou seed whose potential has no $\ell=2$ term at order
$1/r^2$ in its ACMC chart; it would be worthwhile to extend it to the current
octupole $S_3$, whose readout involves $g_{t\phi}$ and hence $\omega'$, and to
the higher mass moments, where the nonlinear terms of \eqref{eq:gttexp} no
longer drop out. Finally, a sharper accounting of the Komar balance
on the rod, where the Maxwell potential is not constant, remains open.

\section*{Acknowledgments}

HMS is supported by LPPM-UNPAR.  AMG was supported by the Natural Sciences and Engineering Research Council of Canada.

\appendix
\numberwithin{equation}{section}

\section{Derivation of the quadrupole scaling law \eqref{eq:M2law}}
\label{app:M2}

This appendix records the definition of the mass quadrupole used in
Section~\ref{subsec:charges} and derives the scaling law \eqref{eq:M2law}.
The argument has two parts: a structural reduction showing that the mass
multipoles of the charged solution depend on the boost only through the
scalar $H$, and an explicit evaluation of the quadrupole in a fixed radial
gauge that yields the closed form $M_2(\alpha)=\tfrac12(1+\cosh^2\!\alpha)
M_2^{(0)}$.

\subsection{The ACMC quadrupole and its invariance}
\label{app:acmc}

For a stationary, axisymmetric, asymptotically flat \emph{vacuum} metric the
Geroch--Hansen mass multipoles \cite{Geroch:1970cd,Hansen:1974zz} coincide
with the coefficients of the asymptotically Cartesian, mass-centred (ACMC)
expansion of Thorne \cite{Thorne:1980ru}. Mayerson \cite{Mayerson:2022ekj}
has extended the Geroch--Hansen construction to stationary non-vacuum
spacetimes by means of an improved twist vector, and has shown that, under
his hypotheses, the equivalence with Thorne's formalism persists: the
coefficients read off in an ACMC chart are the invariant generalized
Geroch--Hansen moments. His hypotheses are (a) sufficient smoothness of the
metric at spatial infinity, (b) the existence of an ACMC chart, and (c) a
topological condition: on the three-dimensional space of orbits of
$\xi=\partial_t$, the closed two-form $\mathbf W^{(2)}=i_\xi{\star}\mathbf V$,
$V^\mu=\bar h^\mu{}_\sigma T^{\sigma\nu}\xi_\nu$ (with $\bar h$ the projector
orthogonal to $\xi$), must be exact --- physically, there may be no net flux of
momentum perpendicular to any spatial two-cycle.

Condition (c) can be proved here. Every spatial two-cycle of the exterior is
homologous to a combination of surfaces of revolution --- a curve in the
$(x,y)$ plane times the $\phi$ circle --- and on such a surface the pull-back
of $\mathbf W^{(2)}$ involves only the components $V^x=T^x{}_t$ and
$V^y=T^y{}_t$ normal to it. For any circular stationary axisymmetric EMD
configuration (Weyl--Papapetrou metric, $A=A_t\,dt+A_\phi\,d\phi$, $\Phi$, all
depending on $(x,y)$ only) both vanish identically: $F_{xy}=0$ and
$\partial_t\Phi=0$, so the momentum density is purely azimuthal. We have
verified $T^x{}_t=T^y{}_t=0$ symbolically for arbitrary functions. Hence
$\mathbf W^{(2)}$ has zero flux through every two-cycle, in particular through
the sphere at infinity, and the improved twist potential exists. As to (a),
the charged metric functions are rational in the prolate coordinates and
analytic functions of $f$ near $f=1$ (through $H$), so their large-$x$
expansions have exactly the structure of those of the Kerr metric written in
the same coordinates, to which Mayerson's theorem applies; we assume, as for
Kerr, that this provides the smoothness required. Condition (b) is established
below, and the task is then to read off the $P_2/R^3$ coefficient of $g_{tt}$
in the ACMC chart. In such a chart, with radial coordinate $R$ and polar angle
$\Theta$ ($w\equiv\cos\Theta$),
\be \label{eq:acmcgtt}
-g_{tt}=1-\frac{2M}{R}
+\sum_{\ell\ge2}\frac{1}{R^{\ell+1}}
\Bigl[c_\ell\,M_\ell\,P_\ell(w)+(\text{lower }P_{\ell'<\ell})\Bigr]
+\dots,
\qquad c_2=-2,
\ee
the $\ell=1$ term being absent because the frame is mass-centered. Thorne's
conditions restrict the \emph{Cartesian} components of the metric: at order
$1/R^k$ they may contain angular harmonics of order at most $k-1$ in $g_{ij}$
(and the analogous restrictions hold for $g_{tt}$ and $g_{ti}$). We impose the
stronger, sufficient condition $g_{ij}=c_k\,\delta_{ij}$ at orders $k=1,2$. In
spherical components, writing $g_{ij}=g_{RR}n_in_j+R^{-2}g_{\Theta\Theta}
\Theta_i\Theta_j+(R\sin\Theta)^{-2}g_{\phi\phi}\Phi_i\Phi_j+\dots$, this is
\emph{isotropy}, $g_{RR}=g_{\Theta\Theta}/R^2=g_{\phi\phi}/(R^2\sin^2\!\Theta)$,
at both orders. Removing $P_2$ from each spherical component would not
suffice: an anisotropy $(g_{RR}-g_{\Theta\Theta}/R^2)\,n_in_j$, even with an
angle-independent coefficient, contains the $\ell=2$ harmonic
$n_in_j-\tfrac13\delta_{ij}$. For the quadrupole it suffices to work to order
$1/R^3$ in $g_{tt}$ with the general axisymmetric, reflection-symmetric
transformation from the prolate coordinates of \eqref{eq:prolate} to
$(R,\Theta)$,
\be \label{eq:acmcmap}
\begin{aligned}
\kappa x &= R\Bigl[1+\frac{a_0+a_2P_2(w)}{R}+\frac{b_0+b_2P_2(w)}{R^2}\Bigr]
+\mathcal{O}(R^{-2}),\\
y &= w+\frac{d_1\,w(1-w^2)}{R}+\frac{d_2\,w(1-w^2)}{R^2}+\mathcal{O}(R^{-3}),
\end{aligned}
\ee
which contains the isotropic and angle-dependent radial shifts \emph{and} the
angular corrections allowed at this order. The ACMC conditions are then
\be \label{eq:gaugecond}
\begin{aligned}
&g_{tt}:\quad \bigl[-g_{tt}\bigr]_{P_2}=0\ \text{at }\mathcal{O}(R^{-2});\\
&\text{spatial, }\mathcal{O}(R^{-1})\text{ and }\mathcal{O}(R^{-2}):\quad
g_{RR}=\frac{g_{\Theta\Theta}}{R^2}=\frac{g_{\phi\phi}}{R^2\sin^2\!\Theta}\
\text{independent of }\Theta,\quad g_{R\Theta}=0;\\
&g_{t\phi}:\quad \ell=1\ \text{only, at }\mathcal{O}(R^{-1})\text{ and }\mathcal{O}(R^{-2}),
\end{aligned}
\ee
and the quadrupole is read from the residual $P_2$ harmonic at the next
order,
\be \label{eq:M2read}
M_2=-\tfrac12\,\bigl[-g_{tt}\bigr]_{1/R^3,\,P_2}.
\ee
For $g_{\phi\phi}$ we use the exact relation \eqref{eq:gijrel}; its second
term, $-(\sinh^2\!\alpha/H)\tilde g_{t\phi}^{\,2}$, is $\mathcal{O}(R^{-4})$
relative to $R^2\sin^2\!\Theta$ (in Cartesian components $\tilde g_{ti}\tilde
g_{tj}=\mathcal{O}(R^{-4})$) and does not enter these conditions.

We have imposed all of the conditions \eqref{eq:gaugecond}, for the vacuum
Kerr and TS$_2$ metrics and for their charged versions, with \emph{both} $p$
and $\beta\equiv\sinh^2\!\alpha$ kept symbolic, so that the following
statements are identities in $(p,\beta)$ on the whole family $0<p<1$,
$\alpha\in\mathbb{R}$, and not checks at sample points. In units $m=1$ the
conditions reduce to the following explicit equations (TS$_2$, $\kappa=p/2$;
for the Kerr seed, $\kappa=p$, replace $p^2$ by $4p^2$):
\be \label{eq:acmcexplicit}
\begin{aligned}
&\bigl[-g_{tt}\bigr]_{1/R^2,\,P_2}=(\beta+2)\,a_2=0
\quad\Rightarrow\quad a_2=0;\\
&\mathcal{O}(1/R)\ \text{spatial}:\quad d_1=0,\quad a_0+d_1=0
\quad\Rightarrow\quad a_0=d_1=0;\\
&\mathcal{O}(1/R^2)\ \text{spatial}:\quad 12b_2+p^2-4=0\ \ (\Theta\text{-dependence of }g_{RR}),\\
&\phantom{\mathcal{O}(1/R^2)\ \text{spatial}:\quad}3b_2-2d_2=0\ \ (g_{R\Theta}),\qquad
8d_2+p^2-4=0\ \ (g_{\Theta\Theta}R^{-2}=g_{\phi\phi}R^{-2}\sin^{-2}\!\Theta),\\
&\phantom{\mathcal{O}(1/R^2)\ \text{spatial}:\quad}16b_0-8b_2+8d_2-p^2=0\ \
(g_{RR}=g_{\Theta\Theta}R^{-2})\\
&\quad\Rightarrow\quad
b_0=\frac{p^2-1}{12}=\frac{4\kappa^2-m^2}{12},\qquad
b_2=\frac{4-p^2}{12}=\frac{m^2-\kappa^2}{3},\qquad
d_2=\frac{4-p^2}{8}=\frac{m^2-\kappa^2}{2}.
\end{aligned}
\ee
The $\mathcal{O}(1/R^2)$ system is overdetermined (four equations for three
unknowns) and consistent: substituting the solution back, every residual
vanishes identically as a rational function of $(p,\beta)$. Three features of
\eqref{eq:acmcexplicit} deserve comment. First, $a_2=0$ because the $1/x^2$
term of $f/H$ carries no $P_2$, and the $g_{t\phi}$ conditions hold
automatically (for TS$_2$ the $\ell=1$ coefficients of $g_{t\phi}/\sin^2\!\Theta$
are $-2q\sqrt{1+\beta}=-2J/m^2$ at $\mathcal{O}(1/R)$ and $2q(1+\beta)^{3/2}$
at $\mathcal{O}(1/R^2)$, with no higher harmonics). Second, the conditions are
\emph{independent of} $\beta$ --- the boost enters the $\mathcal{O}(1/R^2)$
spatial coefficients only through an isotropic shift $\propto\beta^2-2\beta$
common to $g_{RR}$, $g_{\Theta\Theta}/R^2$ and $g_{\phi\phi}/(R^2\sin^2\!\Theta)$
--- so the gauge coefficients $a_2$, $b_0$, $b_2$, $d_2$ are those of the seed:
the ACMC chart of the seed remains ACMC for the charged solution, with
$b_0=(4\kappa^2-m^2)/12$, $b_2=(m^2-\kappa^2)/3$ and $d_2=(m^2-\kappa^2)/2$ for
both seeds (for Kerr, $m^2-\kappa^2=J_0^2/m^2$). Third, the readout \eqref{eq:M2read}
depends on the chart only through $(a_2,b_2)$ --- neither the isotropic shifts
$a_0$, $b_0$ nor the angular corrections $d_1$, $d_2$ enter it --- so in the
computation of Sec.~\ref{app:eval} it is legitimate to retain only $a_2$ and
$b_2$. The chart is therefore ACMC to the required order and the readout is
the invariant quadrupole in the sense of \cite{Mayerson:2022ekj}. As a
calibration, the same chart reproduces the exact Geroch--Hansen values
$M_2=-m^3q^2$ for Kerr and $M_2^{(0)}=-m^3(q^2+p^2/4)$ for TS$_2$,
Eq.~\eqref{eq:seedmoments}.

\subsection{Reduction to the seed potential}
\label{app:reduction}

Write the seed ``Newtonian potential'' $U\equiv 1-f=\mathcal{O}(1/r)$. From
the boost formulae \eqref{eq:KKTSmetric} the two metric components entering
\eqref{eq:acmcgtt}--\eqref{eq:M2read} are
\be \label{eq:reduction}
-g_{tt}=\frac{f}{H}=\frac{1-U}{\sqrt{1+U\sinh^2\!\alpha}},
\qquad
g_{xx}=H\,\tilde g_{xx}=\sqrt{1+U\sinh^2\!\alpha}\;\tilde g_{xx},
\ee
where $\tilde g_{xx}$ is the seed value. Two consequences are immediate and
central. First, both components depend on the seed \emph{only} through $f$
(equivalently $U$) and $\tilde g_{xx}$ (equivalently $e^{2\gamma}$): the
frame-dragging potential $\omega'=\cosh\alpha\,\omega$ does not appear. Since
the ACMC mass moments \eqref{eq:M2read} are built from $g_{tt}$ and $g_{RR}$
alone, they receive no contribution from $\omega'$; the rotational information
that fixes, e.g., the Kerr quadrupole $-m^3q^2$ already resides in $f$ itself.
Second, the entire $\alpha$-dependence is carried by the single scalar
$H=(1+U\sinh^2\!\alpha)^{1/2}$, a function of $U$. Expanding \eqref{eq:reduction},
\be \label{eq:gttexp}
-g_{tt}
=1-\Bigl(1+\tfrac12\sinh^2\!\alpha\Bigr)U
+\tfrac18\sinh^2\!\alpha\,(3\sinh^2\!\alpha+4)\,U^2+\mathcal{O}(U^3),
\ee
and the linear coefficient is exactly
\be \label{eq:linear}
1+\tfrac12\sinh^2\!\alpha=\tfrac12\bigl(1+\cosh^2\!\alpha\bigr)=\frac{M}{m},
\ee
by \eqref{eq:Mcharge}. The scaling law follows from \eqref{eq:linear} by the
following argument, which shows that the nonlinear terms of \eqref{eq:gttexp}
cannot contribute to the quadrupole. In an ACMC chart write
\be \label{eq:Uexp}
U=\frac{u_1}{R}+\frac{u_2(w)}{R^2}+\frac{u_3(w)}{R^3}+\dots,\qquad u_1=2m .
\ee
The ACMC condition $[-g_{tt}]_{1/R^2,P_2}=0$ (no $\ell=2$ harmonic at order
$1/R^2$) reads
$-(1+\tfrac12\beta)[u_2]_{P_2}=0$, i.e.\ $[u_2]_{P_2}=0$ for every $\beta$
(this is the statement $a_2=0$ of Sec.~\ref{app:acmc}). Consequently, for
\emph{any} function $\mathcal{F}(U)$ analytic at $U=0$,
\be \label{eq:nononlinear}
\bigl[\mathcal{F}(U)\bigr]_{1/R^3,P_2}
=\mathcal{F}'(0)\,[u_3]_{P_2}+\mathcal{F}''(0)\,u_1\,[u_2]_{P_2}
=\mathcal{F}'(0)\,[u_3]_{P_2}:
\ee
the $U^2$ term can only feed the $\ell=2$ harmonic at $1/R^3$ through the
product $u_1u_2$, whose $P_2$ part vanishes in an ACMC chart, and
$U^{n\ge3}$ starts at $1/R^3$ with the pure monopole $u_1^3$. Applied to
$\mathcal{F}(U)=(1-U)(1+\beta U)^{-1/2}$ this gives, with \eqref{eq:M2read},
\be \label{eq:M2mech}
M_2(\beta)=\tfrac12\bigl(1+\tfrac12\beta\bigr)[u_3]_{P_2},
\qquad
M_2^{(0)}=\tfrac12[u_3]_{P_2},
\ee
provided $[u_3]_{P_2}$ is evaluated in the same chart for the seed and for the
charged solution. That it may be is the second half of the argument: by
\eqref{eq:gijrel} the spatial components are $g_{ij}=H\tilde g_{ij}$ up to a
term $-(\sinh^2\!\alpha/H)\tilde g_{ti}\tilde g_{tj}=\mathcal{O}(R^{-4})$, with
$H=1+\tfrac12\beta U-\tfrac18\beta^2U^2+\dots$, so their $\mathcal{O}(1/R^2)$
coefficients are those of the seed plus
$\bigl[\tfrac12\beta(u_2+u_1\,2m)-\tfrac18\beta^2u_1^2\bigr]\delta_{ij}$, using
the isotropic $\mathcal{O}(1/R)$ seed coefficient $2m\,\delta_{ij}$; since
$[u_2]_{P_2}=0$, the added terms are proportional to $\delta_{ij}$, the $\mathcal{O}(1/R^2)$ spatial ACMC conditions
--- including isotropy --- are \emph{identical} to those of the seed, the gauge
coefficients $(b_0,b_2,d_2)$ are $\beta$-independent, and the ACMC chart of the
seed is ACMC for the charged solution, exactly as the explicit equations
\eqref{eq:acmcexplicit} show. Therefore
\be \label{eq:M2lawproof}
M_2(\beta)=\bigl(1+\tfrac12\beta\bigr)M_2^{(0)}=\frac{M}{m}\,M_2^{(0)} ,
\ee
for every Weyl--Papapetrou seed with $f=1-2m/r+\mathcal{O}(1/r^2)$ whose
$1/r^2$ term carries no $P_2$ (true for Kerr and TS$_2$, where the
$\mathcal{O}(1/x^2)$ term of $f$ is $2m^2/\kappa^2x^2$). The explicit
evaluation of Sec.~\ref{app:eval} confirms \eqref{eq:M2lawproof} in closed
form and supplies $[u_3]_{P_2}$: in units $m=1$,
$[u_3]_{P_2}=\tfrac43\,(p^2-1)-2b_2$ for both seeds, which with
$b_2=(m^2-\kappa^2)/3$ gives $M_2^{(0)}=-(q^2+p^2/4)$ for TS$_2$ and $-q^2$
for Kerr.

\subsection{Evaluation: exact closed form in $p$ and $\beta$}
\label{app:eval}

Carrying \eqref{eq:reduction} through the gauge conditions
\eqref{eq:gaugecond} and the readout \eqref{eq:M2read} can be done in closed
form, with both the seed parameter $p$ and the boost parameter
$\beta\equiv\sinh^2\!\alpha$ kept symbolic. The only ingredient that makes the
computation tractable is to organize it as truncated series arithmetic rather
than as an expansion of the full rational functions. Writing the chart
\eqref{eq:acmcmap} as $\kappa x=R\,S(t)$ with $t=1/R$ and
$S=1+(a_0+a_2P_2)\,t+(b_0+b_2P_2)\,t^2$, every polynomial $\mathcal{P}(x,y)$ of degree $n$
in $x$ appearing in $f=A/B$ and in $g_{xx}$ becomes
$\mathcal{P}=\sum_k c_k(w)\,\kappa^{-k}S^k\,t^{-k}$, and only the powers $S^k$
truncated at relative order $t^{4}$ are ever needed. Series division then gives
$f$ and $\tilde g_{xx}(\partial x/\partial R)^2$ to the required order, the
boost enters through the finite expansions of $H^{\pm1}$ in $U=1-f$, and the
conditions \eqref{eq:gaugecond} are linear in the gauge coefficients. Every
coefficient is a rational function of $(p,\beta)$, and the final readout
\eqref{eq:M2read} simplifies exactly. For the TS$_2$ seed, in units $m=1$
($\kappa=p/2$),
\be \label{eq:M2beta}
M_2^{\mathrm{TS_2}}(p,\beta)=\frac{(\beta+2)(3p^2-4)}{8}
=-\Bigl(q^2+\frac{p^2}{4}\Bigr)\frac{2+\beta}{2}
=\frac{1+\cosh^2\!\alpha}{2}\,M_2^{(0)},
\ee
and for the Kerr seed ($\kappa=p$),
\be \label{eq:M2betaKerr}
M_2^{\mathrm{Kerr}}(p,\beta)=-\frac{(\beta+2)(1-p^2)}{2}
=-q^2\,\frac{2+\beta}{2}
=\frac{1+\cosh^2\!\alpha}{2}\,M_2^{(0)},
\ee
which is \eqref{eq:M2law} for both seeds, as an identity in $(p,\beta)$, in
agreement with the general argument of Sec.~\ref{app:reduction}: the
$\beta$-dependence of the readout is entirely the linear factor
\eqref{eq:linear}, the nonlinear terms of \eqref{eq:gttexp} contributing
nothing to the $P_2/R^3$ coefficient and the gauge coefficients being those
of the seed. The two limits that
bound the result follow at once: at $\beta=0$ the closed forms return the
Geroch--Hansen quadrupoles \eqref{eq:seedmoments} and $-m^3q^2$, and for the
Kerr seed at arbitrary $\beta$ they reproduce the quadrupole of the boosted
Kerr (Kaluza--Klein) black hole \cite{Rasheed:1995zv,Larsen:1999pp}. The
$\delta=2$ excess $\Delta M_2=-\tfrac14 p^2 m^2M$ of \eqref{eq:DM2law} is the
difference of the two closed forms.

A remark on scope: the twist-based electrovac moment formalism of
\cite{Fodor:2021em}, applied to the charged magnetized TS$_2$ solution of
Einstein--Maxwell theory, would require a dilatonic extension to be applied
here; the ACMC route via \cite{Mayerson:2022ekj} sidesteps this and is the
one we use.

\section*{Declaration of generative AI and AI-assisted technologies in the manuscript preparation process}

During the preparation of this work the authors used Claude (Anthropic) Opus 5.5 in
order to assist with checking the internal
consistency of the derivations and of the cited literature, and writing the codes used for the symbolic checks and to
produce the figures. After using this tool, the authors reviewed and verified the results independently,
and take full responsibility for the content of the published article.


\end{document}